\documentclass[%
reprint,
 amsmath,amssymb,
 aps,
floatfix,
longbibliography,
]{revtex4-1}

\usepackage{graphicx}
\usepackage{grffile} 
\usepackage{subfigure}
\usepackage[marginal]{footmisc}
\usepackage{epstopdf}
\usepackage{amssymb}
\usepackage{mathrsfs}
\usepackage{amsmath,mathtools,amsfonts,bm,graphicx}
\usepackage{upgreek}
\usepackage{color}
\usepackage{etoolbox}
\usepackage[colorlinks=true,linkcolor=blue,urlcolor=blue,citecolor=blue]{hyperref} 
\usepackage{multirow}
\usepackage[normalem]{ulem}

\newcommand{\app}[1]{Appendix~\ref{#1}}

\newcommand{\fig}[1]{Fig.~\ref{#1}}
\newcommand{\Fig}[1]{Figure~\ref{#1}}
\newcommand{\eq}[1]{Eq.~(\ref{#1})}

\newcommand{\Eq}[1]{Equation~(\ref{#1})}

\newcommand{\roundbk}[1]{\left( #1 \right)}

\newcommand{\squarebk}[1]{\left[ #1 \right]}

\newcommand{\bracebk}[1]{\left\lbrace #1 \right\rbrace}

\newcommand{\x}{\mathopen{\textnormal{\textquotesingle}}x\mathopen{\textnormal{\textquotesingle}}}
\newcommand{\y}{\mathopen{\textnormal{\textquotesingle}}y\mathopen{\textnormal{\textquotesingle}}}

\begin{document}

\title{Kauzmann paradox: A possible crossover due to diminishing local excitations}

\author{Xin-Yuan Gao$^1$$^{,}$$^2$}
\author{Chin-Yuan Ong$^2$}
\author{Chun-Shing Lee$^2$}
\author{Cho-Tung Yip$^3$}
\email[Email: ]{h0260416@hit.edu.cn}
\author{Hai-Yao Deng$^4$}
\email[Email: ]{dengh4@cardiff.ac.uk}
\author{Chi-Hang Lam$^2$}
\email[Email: ]{C.H.Lam@polyu.edu.hk}
\address{
$^1$Department of Physics, The Chinese University of Hong Kong, Shatin, New Territories, Hong Kong, China\\
$^2$Department of Applied Physics, Hong Kong Polytechnic University, Hong Kong, China\\		
$^3$Department of Physics, Harbin Institute of Technology, Shenzhen 518055, China\\
$^4$School of Physics and Astronomy, Cardiff University, 5 The Parade, Cardiff CF24 3AA, Wales, UK
}

\date{\today}

\begin{abstract}
  The configurational entropy of supercooled liquids extrapolates to zero at the Kauzmann temperature, causing a crisis called the Kauzmann paradox. Here, using a class of multicomponent lattice glass models, we study a resolution of the paradox characterized by a sudden but smooth turn in the entropy as temperature goes sufficiently low. A scalar variant of the models reproduces the Kauzmann paradox with thermodynamic properties at very low temperatures dominated by correlations. An exactly solvable vector variant without correlation illustrates that a sudden entropy turn occurs when discrete local excitations are largely suppressed. Despite being disordered and infinitely degenerate, the ground states have zero entropy per particle.
\end{abstract}

\maketitle

Despite decades of study, thermodynamic properties of glass formers at deep supercooling remain surprisingly controversial with challenges due to the long relaxation time \cite{stillinger2013,biroli2013,arceri2022}.  
As temperature decreases, the entropy of a supercooled liquid in general drops faster than that of the crystalline counterpart. Simple extrapolations show that the excess entropy of the liquid over the crystal appears to vanish abruptly at a finite temperature called the Kauzmann temperature $T_K$, leading to the so-called Kauzmann paradox~\cite{kauzmann1948}. An early theory by Gibbs and DiMarzio~\cite{gibbs1958} analyzing a lattice model of polymer suggests a second-order phase transition at $T_K$ below which the system becomes 
an ideal glass with zero entropy.
Later, studying excess entropy, Adam and Gibbs~\cite{adam1965} dervied the Vogel-Fulcher-Tamman(VFT) law~\cite{berthier2011}, which is based on a finite $T_K$. This concept forms the cornerstone of major mean-field thermodynamic theories of glass \cite{arceri2022}. Notably, a more modern view proposes that the ideal glass transition is described by a random first-order transition between local glassy clusters and liquid phase~\cite{bouchaud2004}, from which the VFT law or related possible forms can be obtained.
An alternative view is a relatively sudden but smooth crossover with the excess entropy vanishing only at zero temperature~\cite{stillinger1988,wolfgardt1996, johari2000,jack2016,chen2021}. 
Recently, significant progress in experiments~\cite{zhao2013,monnier2021} and molecular dynamics (MD) simulations~\cite{berthier2019, moid2021} was made. In particular, $T_k$ is proposed to be finite in three dimensions but it may be vanishing in two dimensions \cite{berthier2019,moid2021}. However, since a direct and sufficient supercooling for settling the dispute seems unattainable, the paradox remains an actively debated problem \cite{mckenna2017,schmelzer2018}.

Lattice models in general are more efficient computationally than MD simulations by orders of magnitude and relate to theories more readily \cite{garrahan2011,arceri2022}.
However, most are either energetically trivial or of mean-field type and cannot provide detailed answers to the paradox. 
We have recently proposed a distinguishable particle lattice model (DPLM) of glass \cite{zhang2017}, the physical relevance of which has been supported by the observation of many  glassy behaviors
\cite{zhang2017,lulli2020,lee2020,lulli2021,lee2021,gao2022}.
Its thermodynamics properties are exactly solvable.
However, neither can it be applied to the paradox because it assumes infinitely many particle types, leading to a diverging specific entropy.

In this work, we illustrate a simple resolution of the Kauzmann paradox by proposing a class of multi-component lattice models (MCLM), which generalizes the DPLM to a finite number of particle types. They inherit glassy properties from the DPLM but have a finite specific entropy. In particular, a vector variant of the model admits exact solvability at arbitrary temperatures, allowing a thorough and intuitive understanding. 
Below, we explain our simulation and analytic results while more details and verification of their glassy nature are given in the Appendix.

\section{Scalar MCLM} 
\begin{figure}[tpb]
\centering
\includegraphics[width=0.4\textwidth]{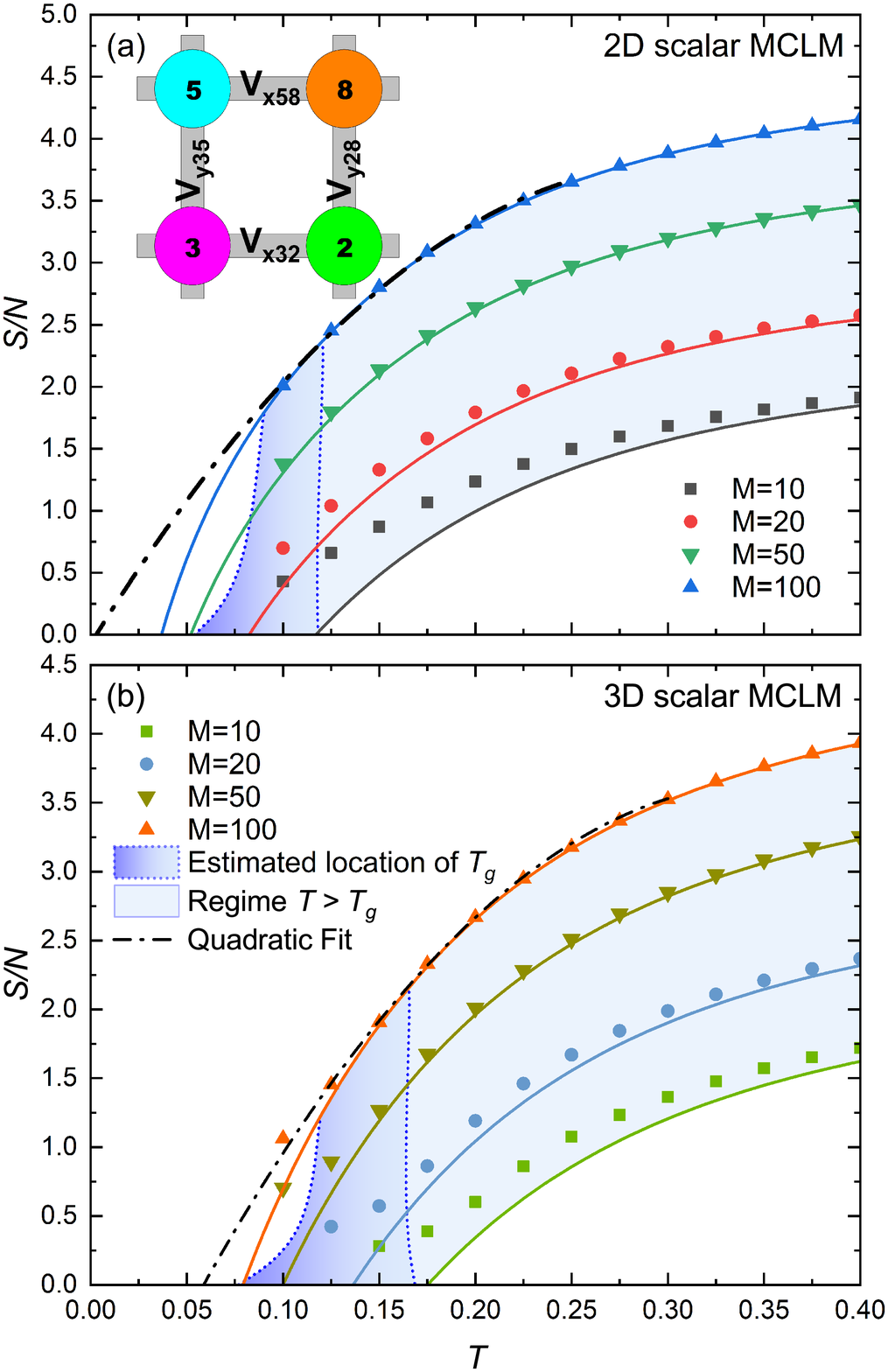}
\caption{Specific entropy $S/N$ against temperature $T$ for scalar MCLM with $M$ particle types in 2D (a) and 3D (b). Simulation results (symbols) agree well with annealed-averaging estimates (solid lines) at large $M$ and $T$. As $T$ decreases, $S/N$ first appears to drop to 0 at a finite $T$, exemplifying the Kauzmann paradox. For $M\le 50$ in 3D, the drop slows down at low $T$ signifying the onset of smooth turns. For $M=100$, a simple quadratic function (dashed-dotted lines) extrapolates $S/N$ to 0 ( $> 0$ ) in 2D (3D). Possible values of $T_g$ for various $M$ are indicated by a region shaded in dark gradient blue, while the light blue region indicates $T > T_g$.
Inset in (a): An example configuration in a $2\times 2$ region showing particle types (numbers in circles) and associated pair interactions $V_{\alpha_{ij} s_i s_j }$. 
}
\label{Vakl}
\end{figure}

We first define a scalar version of the model in 2D and it is similar in 3D. A square lattice of width $L$ following periodic boundary conditions is occupied by $N \le L^2$ particles. No two particles can sit at the same site. Particles fall into $M$  types, i.e. $M$ components, and there are $N/M$ particles for each type.  
Nearest neighbors of types $k$ and $l = 1, 2, \dots, M$ interact with energy $V_{\alpha k l}$, where $\alpha=\x$ or $\y$ if particle $l$ is to the East or North respectively of particle $k$ (see inset in  \fig{Vakl}(a)).
In general, $V_{\alpha k l}$ differs from  $V_{\alpha l k}$. As a consequence, for a $d$-dimensional MCLM with $M$ types, $V_{\alpha k l}$ totally takes $M^2d$ independent discrete values. 

At the beginning of a simulation, the whole set of $V_{\alpha k l}$ are randomly sampled from a distribution $g(V)$ which is taken as the uniform distribution in $[0,\Delta V]$. We take dimensionless units by putting $\Delta V=1$ and the Boltzmann constant $k_B=1$. Possible particle segregation and crystallization into domains of a single type or of two types in checkerboard arrangements are suppressed by the anisotropic form of the interaction $V_{\alpha k l}$.
Physically, the anisotropy effectively accounts for different frustration at different sites and directions due to further neighbors in the disordered configurations. For large $M$ or after ensemble averaging, system isotropy is restored. 
Let $s_i$ be the particle type at site $i$. The total energy of the system is
\begin{equation}
E=\sum_{\left \langle i,j \right \rangle ^*} V_{\alpha_{ij} s_i s_j},
\label{systemEnergy}
\end{equation}
where the sum is overall occupied nearest-neighbor sites $i$ and $j$, with the asterisk denoting $j$ at the East or North of $i$ corresponding to $\alpha_{ij}=\x$ or $\y$ respectively.
The MCLM reduces to the DPLM when $M$ is large. 

We conduct equilibrium simulations on fully occupied lattices in 2D and 3D adopting highly efficient non-local swap dynamics \cite{ninarello2017,gopinath2022}.
In each Monte Carlo step, two particles in the system are randomly chosen. They are swapped based on the Metropolis algorithm at temperature $T$ by an acceptance probability
\begin{equation}
p=\left\{\begin{array}{ll}
\exp(-\Delta E/k_BT) ~~~~~ &\text{if ~} \Delta E\ge 0\\
1& \text{otherwise} \\
\end{array}\right. 
\end{equation}
where $\Delta E$ is the change of the system energy $E$ due to the swap and $k_B=1$ is the Boltzmann constant.
We measure $E$ at various temperatures $T$ up to $T_1=1$, at which high-temperature analytic solutions are accurate using the annealed averaging approximation explained in \app{Sec:annealed}. The system entropy $S$, which equals the configurational entropy in the absence of vibration, is then computed as
\begin{equation}
  \frac SN=\frac {S_1}N -  \int_{T}^{T_1}\frac{c_V}{T} \mathrm{d}T.
\end{equation}
Here, $c_V$ is the specific heat measured at $T$.
\Fig{Vakl} plots the obtained specific entropy $S/N$ against temperature. The excellent efficiency of the swap algorithm allows the study even close to the glass transition temperature $T_g$, estimated by extrapolating kinetic simulation data to experimental time scales following \cite{ninarello2017}. The results on the entropy qualitatively resemble those in experiments and in general extrapolate to zero at a finite temperature, reproducing the Kauzmann paradox. For $M= 10$, 20 and 50 in 3D, we observe smooth turns of the entropy at $T \simeq 0.17$, 0.15 and 0.12 and respectively close to $T_g$, avoiding the entropy crisis. However, difficulty in measurement at very low temperatures forbids us to determine if such turns occur also for larger $M$ and in 2D.

Concerning analytical treatments, a remarkable feature of the DPLM is that its thermodynamic properties are exactly solvable at all $T$ and can be calculated based on simple annealed averages \cite{zhang2017}, a result which has been extensively verified in simulations \cite{zhang2017,lulli2020,lee2020}. The MCLM inherits this property, but only approximately.  \Fig{Vakl} shows $S/N$ from annealed averaging calculations, which agrees well with simulations for large $M$ and $T$. In particular, the approximation exhibits the Kauzmann paradox, reaching a zero entropy at the Kauzmann temperature $T_K$ given by (see \app{section::noexcitation})
\begin{equation} 
  T_K = \frac{\Delta V}{e k_B M^{1/d}},
  \label{TK1}
\end{equation}
in $d$ dimensions. Note that \eq{TK1} is derived for a uniform interaction distribution $g(V)$ modeling a strong glass \cite{lee2020} and results differ quantitatively for other choices. From both simulation results in \fig{Vakl} and \eq{TK1}, we observed that $T_K$ is smaller in 2D than in 3D for any given $M$.  Moreover, applying a quadratic extrapolation~\cite{elmatad2010} for $M=100$ (dashed lines in \fig{Vakl}), we get $T_K\simeq 0$ in 2D but $T_K>0$ in 3D. This may explain a suggested qualitative difference between dimensions motivated by recent MD simulations ~\cite{berthier2019, moid2021,guiselin2022}.

\begin{figure}[tb]
\centering
\includegraphics[width=0.4\textwidth]{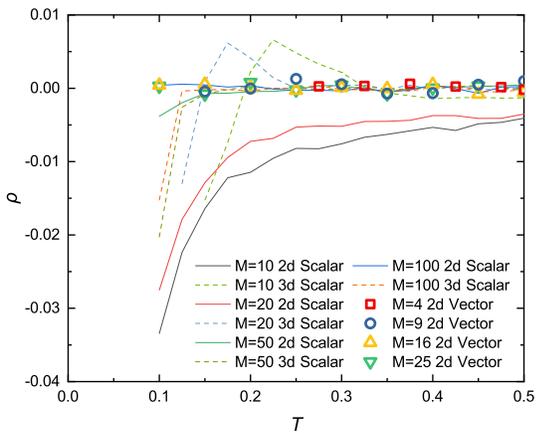}
\caption{Nearest-neighbor bond-bond correlation $\rho$ against $T$ for scalar and vector MCLM with $M$ particle types. }
\label{Corr}
\end{figure}

The annealed averaging approximation effectively assumes statistical independence of all pair interactions. To examine this, we measure a normalized interaction correlation defined by $\rho= {\mathrm{cov}(V_1,V_2)}/{\mathrm{var}(V_1)}$. Here, '$\mathrm{cov}$' and '$\mathrm{var}$' denote covariance and variance, respectively, and $V_1$ and $V_2$ are neighboring pair interactions in perpendicular directions sharing one common particle. 
From results shown in \fig{Corr}, $\rho$ is noticeably non-zero mainly for $M \alt 20$ at $T \alt 0.3$. This regime coincides with that in which the measured entropy deviates significantly from the annealed averaging approximation as observed in \fig{Vakl}(a). In particular, considerable correlations are present at small $S/N \alt 1$. Therefore, the zero or negative entropy predicted by \eq{TK1} occurs when the approximation should break down.
An improved approximation gives instead a positive entropy at all finite temperatures  (see \app{section:uncoptcle}). Such correlations, bound to exist also in realistic systems, present the main difficulty preventing a thorough understanding of the paradox and in particular a possible turn in the entropy. 

\section{Vector MCLM}
We now introduce a vector MCLM which suppresses both crystallization and correlation, allowing exact thermodynamics. 
A motivation for this variant of the MCLM originates from \eq{TK1}, where one sees that $M$ and $d$ affect $T_K$ via the quantity $\mathcal{M}=M^{1/d}$. 
Since $\mathcal{M}$ can be interpreted as the number of types per dimension, we attempt to denote a particle type in the $d$-dimensional model by a $d$-dimensional vector $\mathbf{k}=(k(x),k(y),k(z),\dots)$, where all type indices $k$ takes $k=1, 2, \dots,\mathcal{M}$ so that the number of particle types is $M=\mathcal{M}^d$. In the following discussion, we focus on 2D simulation and show that our exact predictions are highly accurate. Results in higher dimensions are expected to behave similarly: It can be shown that all thermodynamical quantities differ from the 2D results by a factor $d/2$(see \app{section::exactthermodynamics}). 

\begin{figure}[htpb]
\centering
\includegraphics[width=0.4\textwidth]{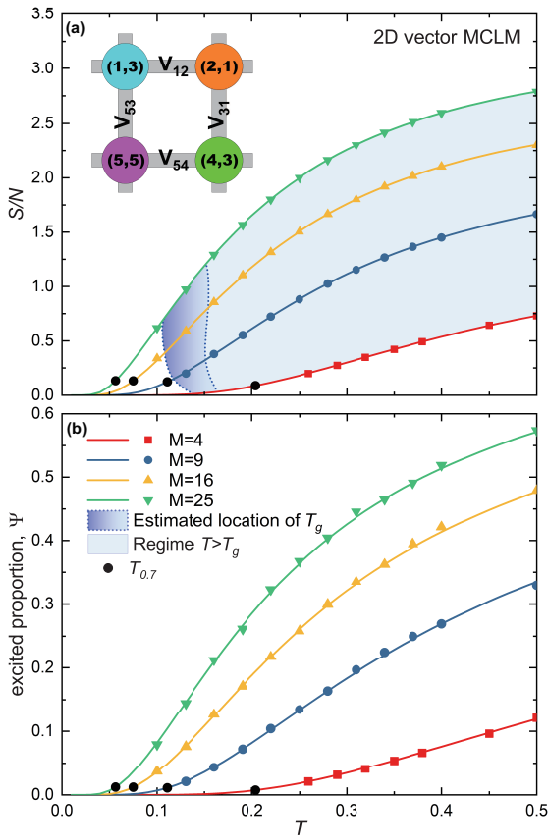}
\caption{(a) Comparison of 2D vector MCLM simulation results (symbols) with exact analytic calculations (solid lines) for entropy per particle $S/N$ and (b)  
the proportion of excited pair interactions plotted against temperature $T$. The drop of $S/N$ slows down at $T_{0.7}$ (black symbols in (a) and (b)) when the slope has reduced to 70\% of its maximum value. The blue shading in (a) is defined similarly to that in \fig{Vakl}.
Inset in (a): An example configuration in a $2\times 2$ region showing particle types $\mathbf{s_i}=(s_{i}(x),s_{i}(y))$
and pair interactions $V_{s_{i}(\alpha_{ij}) s_{j}(\alpha_{ij})}$.
}
\label{exact}
\end{figure}

Nearest neighbors of types $\mathbf{k}$ and $\mathbf{l}$ interact with energy $V_{k(\alpha)l(\alpha)}$, where $\alpha=\x$ or $\y$ if particle $\mathbf l$ is to the East or North of particle $\mathbf k$ (see inset in \fig{exact}(a)). The interaction at a horizontal (vertical) bond thus depends only on the $x$ ($y$) type index. Let $\mathbf s_i = (s_{i}(x), s_{i}(y))$ be the particle type at site $i$. The system energy is 
\begin{equation}
E=\sum_{\left \langle i,j \right \rangle ^*} V _{s_{i}(\alpha_{ij}) s_{j}(\alpha_{ij}) }
\label{Evector}
\end{equation}
where the sum and $\alpha_{ij}$ are defined similarly to those in \eq{systemEnergy}. To enable exact thermodynamics, we define  deterministically
\begin{equation}
V _{k(\alpha) l(\alpha)}=\frac{[k(\alpha)-l(\alpha)+1]\ \mathrm{mod}\ \mathcal{M}}{\mathcal{M}-1} ~ \Delta V
\label{Vkala}
\end{equation}
with $\Delta V=1$. All particle types are then similar to each other because they follow a cyclic symmetry in which the system is invariant under relabelling the types via ${k \rightarrow (k \mathrm{~mod~} \mathcal{M})+1}$.
This definition leads to discrete energies 
\begin{equation}
V _{k(\alpha) l(\alpha)} = 0, ~\delta V, ~2\delta V,~ \dots, ~\Delta V,
\label{Vdiscrete}
\end{equation}
with $\delta V = \Delta V/(\mathcal{M}-1)$. Indeed, $V _{k(\alpha) l(\alpha)}$ follows a uniform discrete  distribution $g(V)$, which approaches the continuous uniform distribution in $[0, \Delta V]$ at large $M$.  

\Fig{exact}(a) shows the specific entropy $S/N$ measured from equilibrium simulations using the procedures explained above. Results resemble those from scalar MCLM, exhibiting the Kauzmann paradox. Exact analytic results (see \app{section::exactthermodynamics}) are also displayed, showing excellent agreement with simulations even at the lowest temperatures studied. Correlation $\rho$ between neighboring interactions is consistent with zero at all temperatures as shown in \fig{Corr}. 

The exact results in \fig{exact}(a) show a smooth turn of the entropy at low temperatures for all $M$.
The cause of the turn can be understood as follows. At high temperatures, the model is a simple lattice gas in which particles can neighbor any other, resulting in high entropy. As temperature decreases, only pairings with interactions of order $k_BT$ or below are energetically favorable and the entropy thus drops. Noting that interactions take discrete values according to \eq{Vdiscrete}, the turn occurs when
almost all excited interactions are suppressed. To show this, \fig{exact}(b) plots the measured excited proportion $\Psi$ of the pair interactions, i.e. those with energies $\ge \delta V$, which again well agrees with the exact results. The proportion $\Psi$ is below 1.5\% at temperature $T_{0.7}$
at which the slope of $S/N$ has decreased to 70\% of its maximum value (black circles).
The system is very close to a ground state with few excitations. A further rapid drop in the entropy is thus impossible and the entropy turn follows.

\begin{figure}[tpb]
\centering
\includegraphics[width=0.4\textwidth]{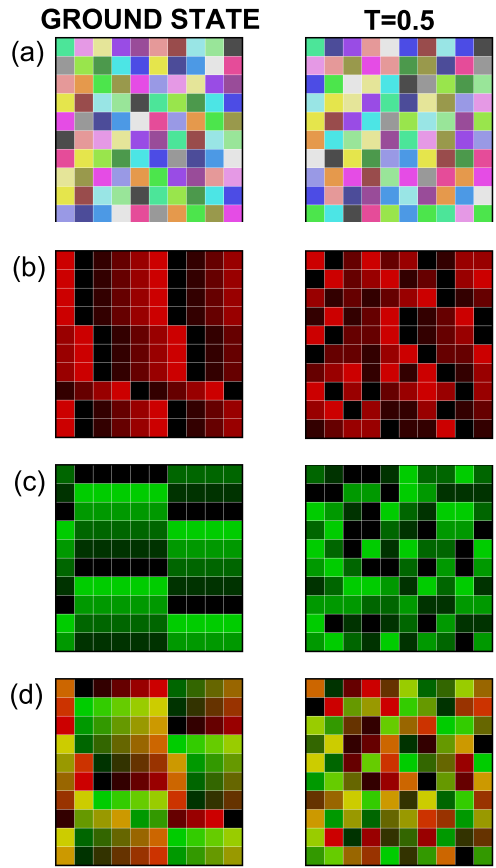}
\caption{
  Particle configurations from small-scale 2D vector MCLM simulations with  $N=100$ particles of $M=25$ types at $T=0$ (left) and 0.5 (right). At $T=0$, system energy $E=0$, implies a ground state. Particle type $\mathbf{s_i}=(s_{i}(x),s_{i}(y))$ at site $i$ is colored (a) randomly for each value of $\mathbf{s_i}$, (b) red with brightness $s_{i}(x)$, (c) green with brightness $s_{i}(y)$, and (d) red and green superimposing those in (b) and (c), i.e. a RGB color code of $\bracebk{s_{i}(x),s_{i}(y), 0}$.
 }
\label{real_space_image}
\end{figure}

The vector MCLM exhibits intriguing ground state properties.
From exact results shown in \fig{exact}, only at $T=0$ does $S/N$ vanish. The system is then in its ground state with energy $E=0$ since all pair interactions equal 0. We obtain a zero-temperature entropy ${S_0=\sqrt{N} k_B \ln M}$  (see \app{section::groundstate}), which diverges at large $N$ while $S_0/N$ vanishes. This implies highly degenerate disordered ground states. \Fig{real_space_image}(a) shows the particle arrangement of a typical ground state in a small system, with each particle type colored randomly  (left panel). It appears remarkably disordered and resembles the high-temperature configuration similarly colored  (right panel). There are however hidden regularities. For example, particles of the same type at the same column or row must be separated by a distance of a multiple of 5. In addition, only five possible particle types can form nearest neighbors in a certain direction of a given type.  To elucidate these features, \fig{real_space_image}(b)-(d) recolor each type with red and/or green color components in the red-green-blue (RGB) color space depending on the type indices $s_{i}(x)$ and $s_{i}(y)$. Strict cyclic order of $s_{i}(x)$ and $s_{i}(y)$ in $x$ and $y$ directions respectively is now clearly revealed. On the other hand, the lack of order of $s_{i}(x)$ and $s_{i}(y)$ respectively in $y$ and $x$ directions explains the large $S_0$ associated with the high ground-state degeneracy.

\section{Discussions}

The Kauzmann paradox is a long-standing problem in the study of glass. The MCLM is a non-mean field microscopic model of glass which illustrates a resolution of the paradox in detail in finite dimensions. Moreover, they are generalizations of the DPLM which have already demonstrated many glassy properties
\cite{zhang2017,lee2020,lulli2020,lulli2021,lee2021,gao2022}. This guarantees that the resolution is consistent with the diverse phenomena of glass, which cannot be achieved if completely different models are invoked to explain different properties.

For the vector MCLM, the sudden turn of the entropy occurs when most pair interactions have reached the lowest discrete level. Then, specific entropy cannot drop significantly further and hence turns. A central premise of the paradox is that there must either be a phase transition or a smooth but surprisingly sudden entropy turn, definitive evidence for neither is yet unavailable.  We have shown here that a sudden turn emerges naturally in the vector MCLM.
In fact, this may be commonplace for systems with discrete local levels, which play a pivotal role in theoretical studies of glass \cite{adam1965}.
More generally, a quantum harmonic oscillator also exhibits a similar entropy turn (see \app{section::harmonicoscillator}).

The annealed averaging approximation for the scalar MCLM predicts a zero entropy at a finite temperature $T_K$ in \eq{TK1},  analogous to an early result of Gibbs and DiMarzio~\cite{gibbs1958}. The approximation is based on the notion of uncorrelated interactions, which however breaks down close to $T_K$, thus invalidating the prediction. In \app{section:uncoptcle}, we derive alternative uncorrelated particle approximations. Without excitation, a similar finite $T_K$ is obtained. After properly including excitations, the entropy then remains positive for all finite temperatures, which we believe to be a more reasonable result. 
Note that Gibbs and DiMarzio's calculation also have been analogously amended \cite{milchev1983,wittmann1991,wolfgardt1996}.

Pair interactions between neighboring particles in inherent structures of glass \cite{stillinger1984} can take many possible values not only because of different particle types but also of momentarily quenched random particle separations due to frustration.
Lattice models of a single or few particle types in contrast provide intrinsically only a few possible interactions.  To remedy this, the DPLM and the MCLM consider many particle types leading to a distribution of interactions, which has been crucial to account for Kovacs paradox \cite{lulli2020}, the heat capacity of two-level systems \cite{gao2022}, etc.
The presence of multiple particle types is directly justifiable for polydisperse and polymer systems. When applied to binary fluids or small molecular glasses, the extra particle types provide a simple approach to effectively model the diverse magnitudes of interactions in the presence of sublattice particle displacements or orientations. Therefore, the parameter $M$ in the MCLM does not only account for the different particle types but also molecular freedoms truncated in lattice models. This may explain why results reported here are more compatible with real glass formers for $M \agt 10$, while glasses may only be of one or two components. 
When applying the MCLM to model crystalline multi-component systems such as high-entropy alloys \cite{yeh2013}, we expect that $M$ can then be identified directly with the actual number of particle types.

The highly degenerate disordered ground states of the vector MCLM with a non-extensive entropy may be a concrete finite-dimensional example of the elusive ideal glass \cite{gibbs1958}, although strictly speaking they only occur without excitation at zero temperature.  They are perfectly degenerate due to the simple form of pair interactions in \eq{Evector}.  Introducing a small spread among the interactions should fine-split the ground states but should not alter our conclusions qualitatively.
Elementary excitations at low temperatures are localized to individual interactions and follow exactly solvable statistics. 
There is no underlying finite-temperature phase transition. Although it may be impossible to rule out a phase transition at some $T_K$ in some or most realistic glasses, our results should provide a direct example in finite dimensions showing how such a transition is not essential in resolving the paradox.

In both scalar and vector MCLM discussed in the present work, discrete energy levels of bonds are assumed. These discrete levels reflect that for any particle configuration representing an inherent structure with vibrations disregarded, local relaxations only take a countable number of possible routes. This is analogous to commonly assumed properties of cooperatively rearranging regions in, for example, Adam-Gibbs theory~\cite{adam1965}. Nevertheless, whether the discrete level assumption applies to realistic glass formers or not is still under intensive debate.

In summary, we have studied scalar and vector variants of the MCLM of glass in order to illustrate in a concrete manner a possible resolution of the Kauzmann paradox in lattice models. Whether this resolution can be applied to realistic glasses is a long-standing open question deserving further investigation.
For the scalar MCLM, an entropy turn, which avoids the entropy crisis, is directly observed under certain conditions. It is found that with the same number of particle types $M$, values $T_K$ in  2D systems are naturally much lower than those in 3D.
For the vector MCLM with correlations suppressed, an exact calculation, well verified by simulations, show that the entropy turn occurs in all cases when almost all pair interactions have reached a discrete lowest energy state. 
The ground states of the vector MCLM are amorphous and infinitely degenerate but possess a vanishing specific entropy.

This work was supported by General Research Fund of Hong Kong (Grant 15303220) and
National Natural Science Foundation of China (Grant 11974297).

\appendix

\tableofcontents

\section{Model details and simulation methods}
\label{model}

For the scalar MCLM, there are $M$ particle types and $s_i$ denotes the type at site $i$. The system energy is
\begin{equation}
E=\sum_{\left \langle i,j \right \rangle ^*} V_{\alpha_{ij} s_i s_j}
\label{app:systemEnergy}
\end{equation}
where the sum is overall occupied nearest neighboring sites $i$ and $j$ assuming, for example, in 2D that site $j$ is at the East (for $\alpha_{ij}=\x)$ or the North (for $\alpha_{ij}=\y$) of site $i$. 
We select before the start of each simulation all pair-interactions $V_{\alpha k l}$ from the $a$ $priori$ distribution $g(V)$, which is taken as the uniform distribution in $[0, \Delta V]$ with $\Delta V=1$. Generalization to 3D is straightforward.

In $d$ dimensions, there are $M^2d$ possible interactions in the system. To avoid large sample-to-sample fluctuations at small $M$, the random sampling is performed by  shuffling the set of equally-spaced discrete values ${V_\mu}$ where
\begin{equation}
  V_\mu=\frac{\mu-1}{M^2d-1} \Delta V
  \label{Vscalar}
\end{equation}
with $\mu=1, 2, \dots, M^2d$,  before randomly assigned to $V_{\alpha k l}$. 

The vector MCLM in 2D considers $M=\mathcal{M}^2$ particle types. The type at site $i$ is denoted by a 2D vector $\mathbf s_i = (s_{i}(x), s_{i}(y))$, where the type indices $s_{i}(x)$ and $s_{i}(y)$ equal $1,2, \dots, \mathcal{M}$. 
The system energy is
\begin{equation}
E=\sum_{\left \langle i,j \right \rangle ^*} V _{s_{i}(\alpha_{ij}) s_{j}(\alpha_{ij}) }
\label{app:Evector}
\end{equation}
where the sum and $\alpha_{ij}$ are defined similarly as above. To enable exact solvability, the interaction $V _{k(\alpha) l(\alpha)}$ is given deterministically by
\begin{equation}
V _{k(\alpha) l(\alpha)}=\frac{[k(\alpha)-l(\alpha)+1]\ \mathrm{mod}\ \mathcal{M}}{\mathcal{M}-1} ~ \Delta V
\label{app:Vkala}
\end{equation}
with $\Delta V=1$. The possible values of the interactions are
\begin{equation}
  V_\mu=\frac{\mu-1}{\mathcal{M}-1} \Delta V
  \label{Vvector}
\end{equation}
with $\mu=1, 2, \dots, \mathcal{M}$.

For both MCLM variants, simulation algorithms are similar.
As discussed in the main text, a Monte Carlo non-local swap algorithm is used both to measure equilibrium properties and to obtain initial equilibrium configurations for kinetic simulations.

To study dynamical properties, after the system has been equilibrated using the swap algorithm, particle motions are further simulated using a void-induced dynamics \cite{zhang2017,lee2020,lulli2020}. We consider a small void density $\phi_v=0.01$. Each particle can hop to a nearest neighboring void under a rate
\begin{equation}
w=\left\{\begin{array}{ll}
w_0\exp(-\Delta E/k_BT) ~~~~~ & \text{if ~} \Delta E\ge 0\\
w_0&\text{otherwise} \\
\end{array}\right.
\end{equation}
where $w_0=10^{-6}$. 

In all our main simulations, we consider lattices of side length $L=100$ in 2D and $L=20$ in 3D.
Results have been verified to be for systems in equilibrium without noticeable finite size effects.

\section{Glassy dynamics and local correlations}

The DPLM exhibits many characteristic features of glass
including stretched relaxations \cite{zhang2017} a wide range of fragilities \cite{lee2020}, Kovacs paradox \cite{lulli2020}, Kovacs effects \cite{lulli2021}, heat capacity overshoot \cite{lee2021}, low-temperature heat capacity of two-level systems \cite{gao2022}, and diffusion coefficient power-laws under partial-swap \cite{gopinath2022}.
Since the MCLM reduces to the DPLM when the number of particle types $M$ is large. These properties are expected to be applicable to the MCLM at sufficiently large $M$. As a verification, we measure here fundamental dynamical quantities following \cite{zhang2017} for all systems studied in the main text. Correlations are also quantitatively measured. 

\subsection{Diffusion coefficient}
\begin{figure}[tb]
\centering
\includegraphics[width=0.4\textwidth]{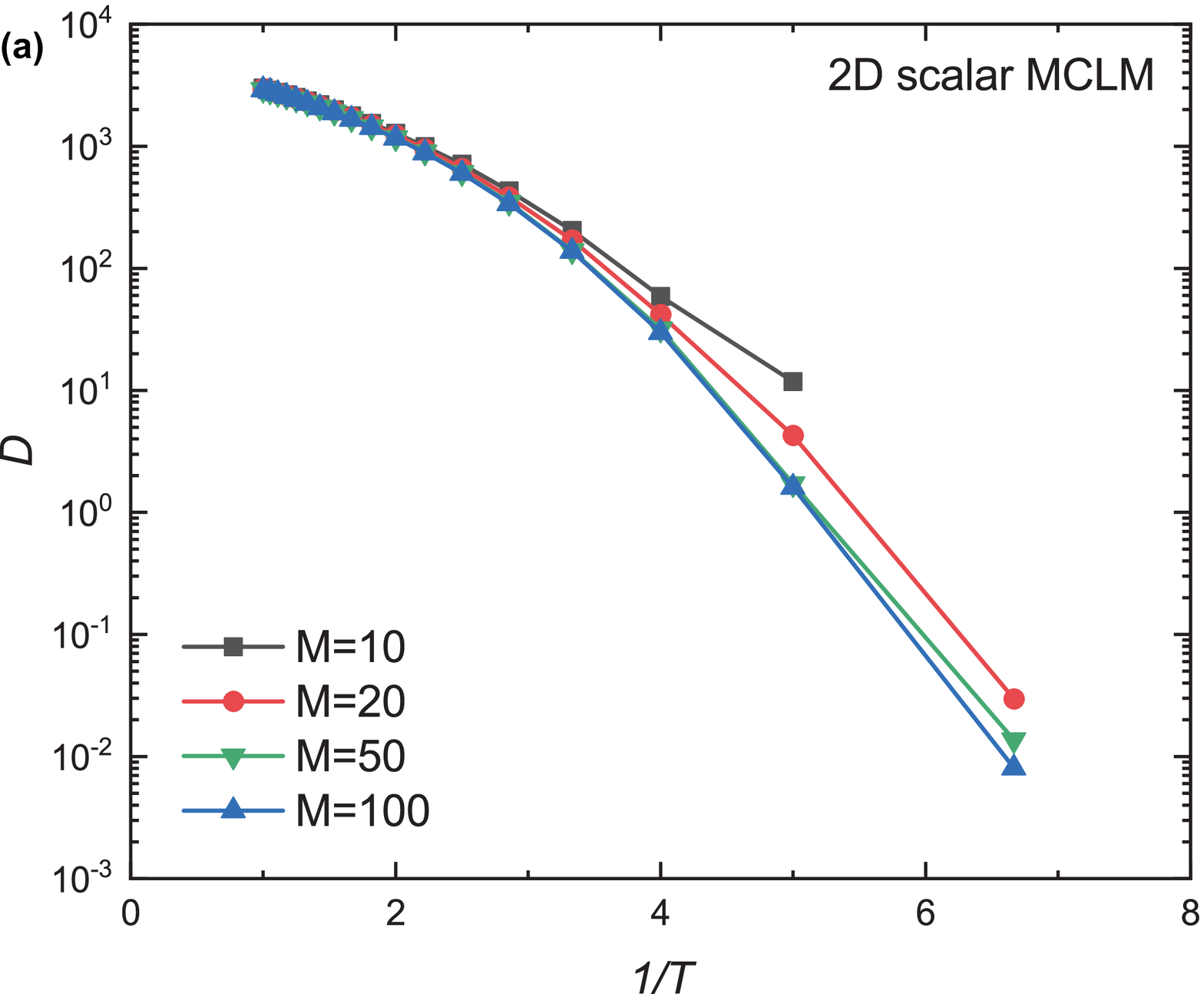}
\includegraphics[width=0.4\textwidth]{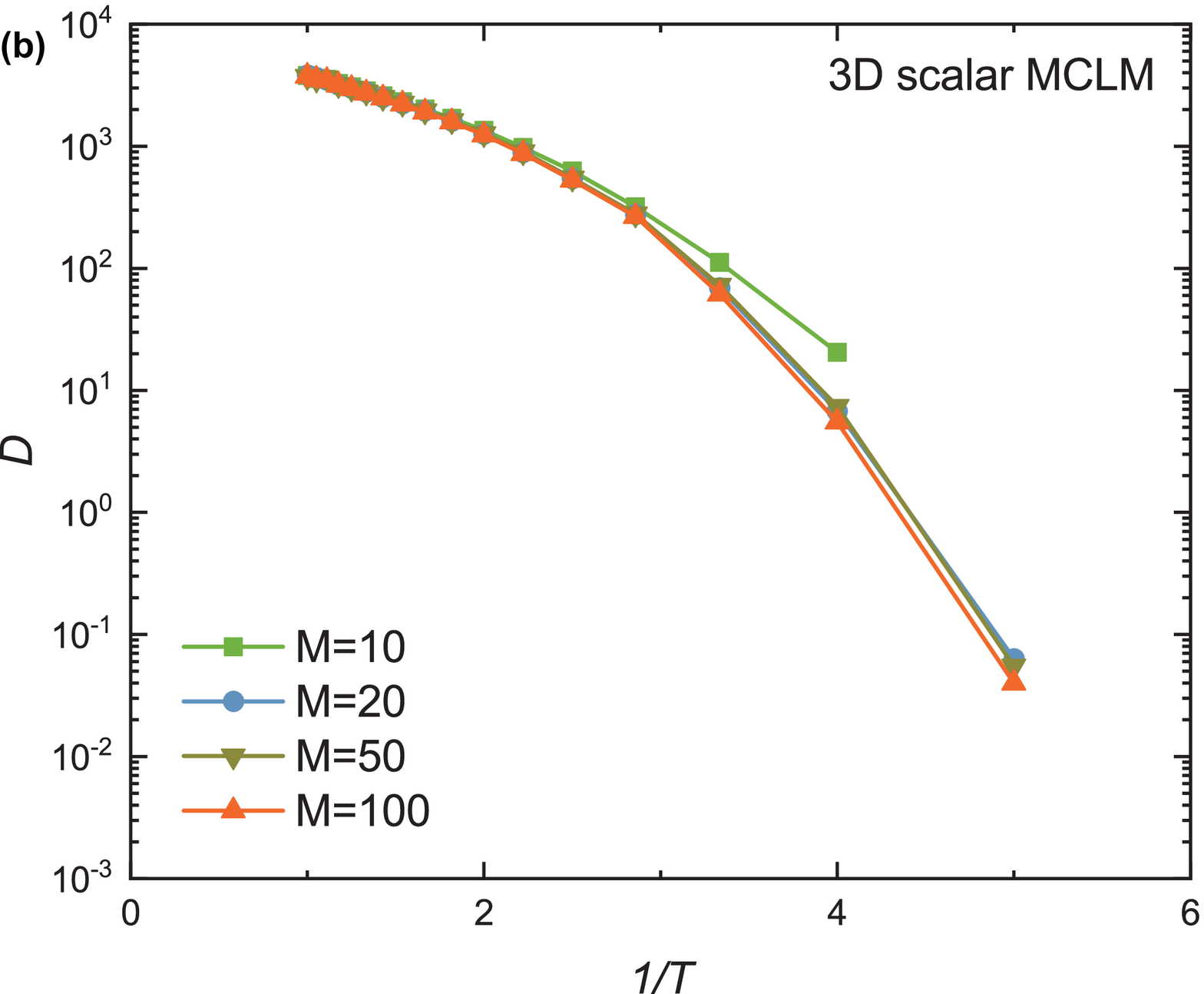}
\includegraphics[width=0.4\textwidth]{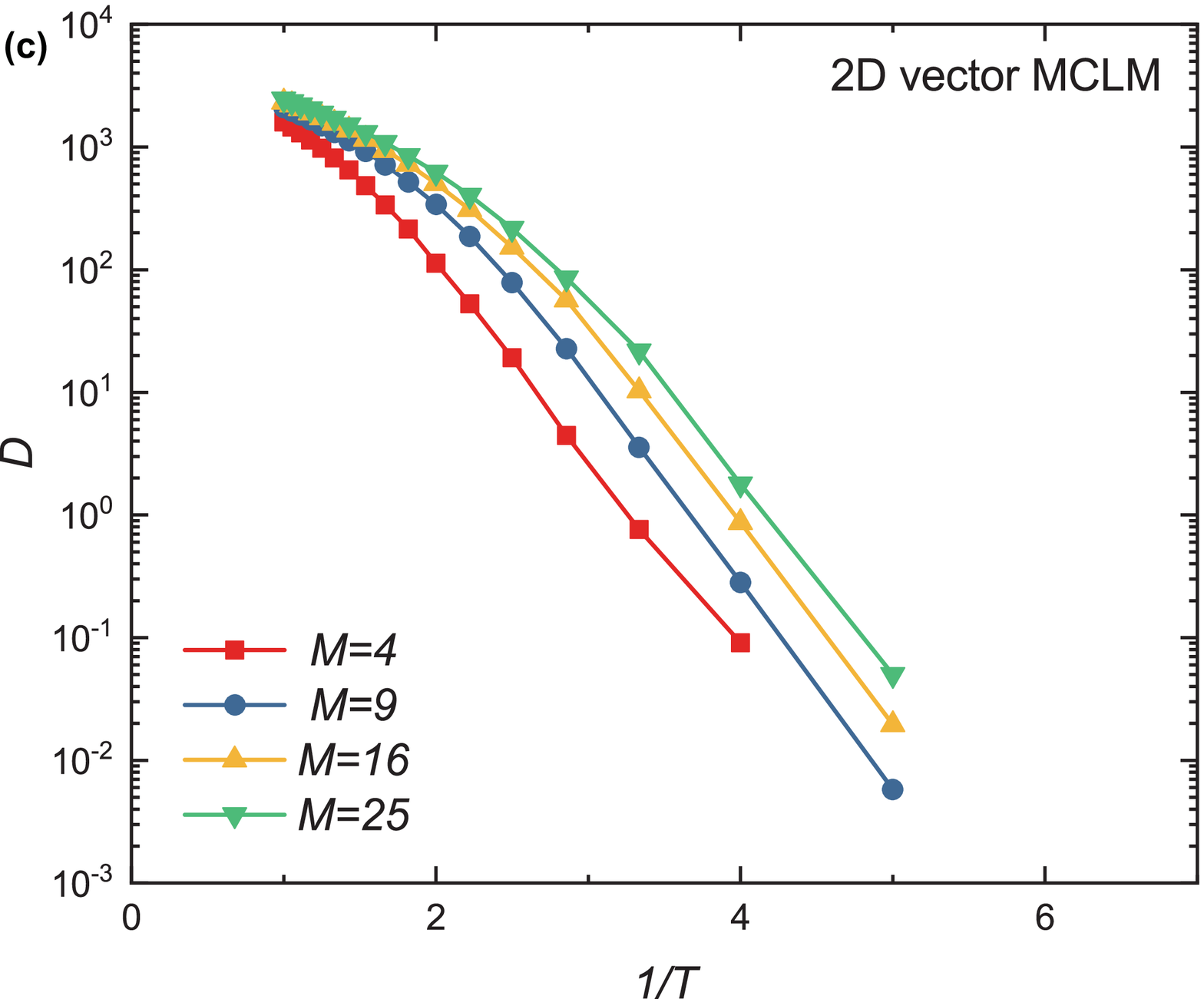}
\caption{Diffusion coefficients $D$ against $1/T$ for 2D scalar (a), 3D scalar (b) and 2D vector (c) MCLM with $M$ particle types.
}
\label{D}
\end{figure}
The particle diffusion coefficient is defined as
\begin{equation}
D=\frac{1}{2d}\lim_{t\rightarrow\infty}\frac{\left\langle\left|\mathbf{r}_l(t)-\mathbf{r}_l(0)\right|^2\right\rangle}{t} 
\end{equation}
where $\mathbf{r}_l(t)$ denotes the position of particle $l$ at time $t$. \Fig{D} plots $D$ against $1/T$ obtained from kinetic simulations. We observe the super-Arrhenius temperature dependence characteristic of glass in all cases.

\subsection{Relaxation time and stretching exponent}
\begin{figure}[tb]
\centering
\includegraphics[width=0.4\textwidth]{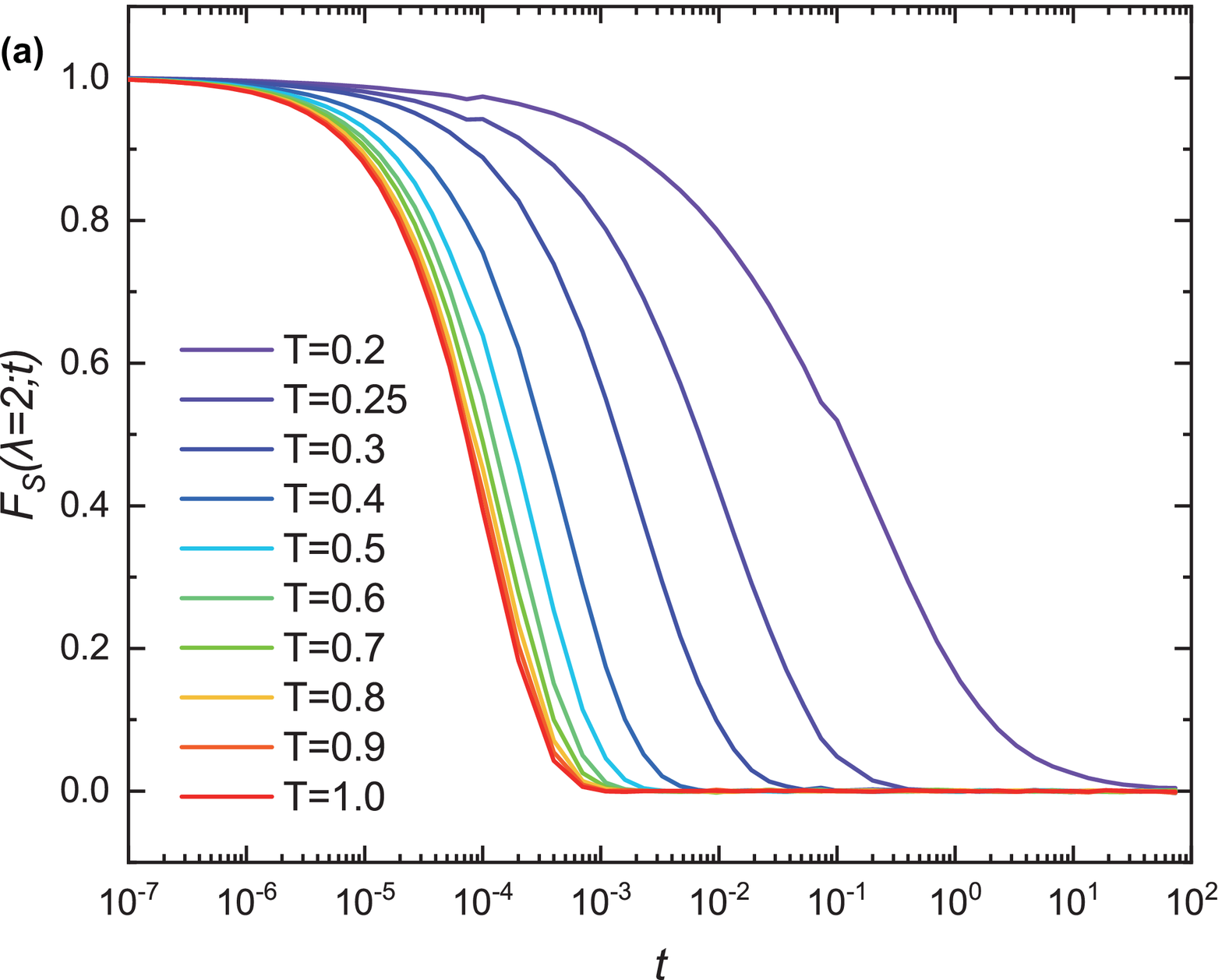}
\includegraphics[width=0.4\textwidth]{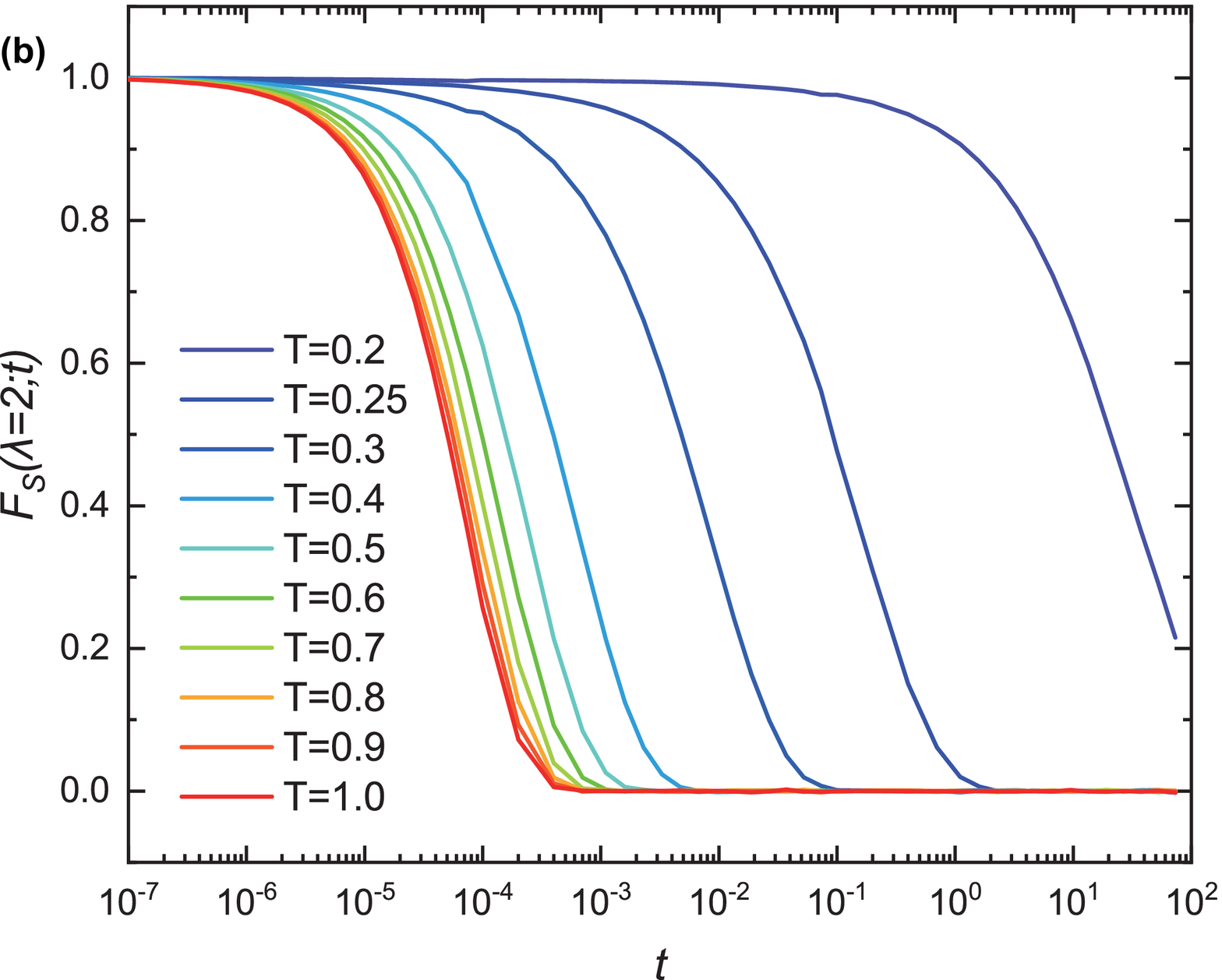}
\includegraphics[width=0.4\textwidth]{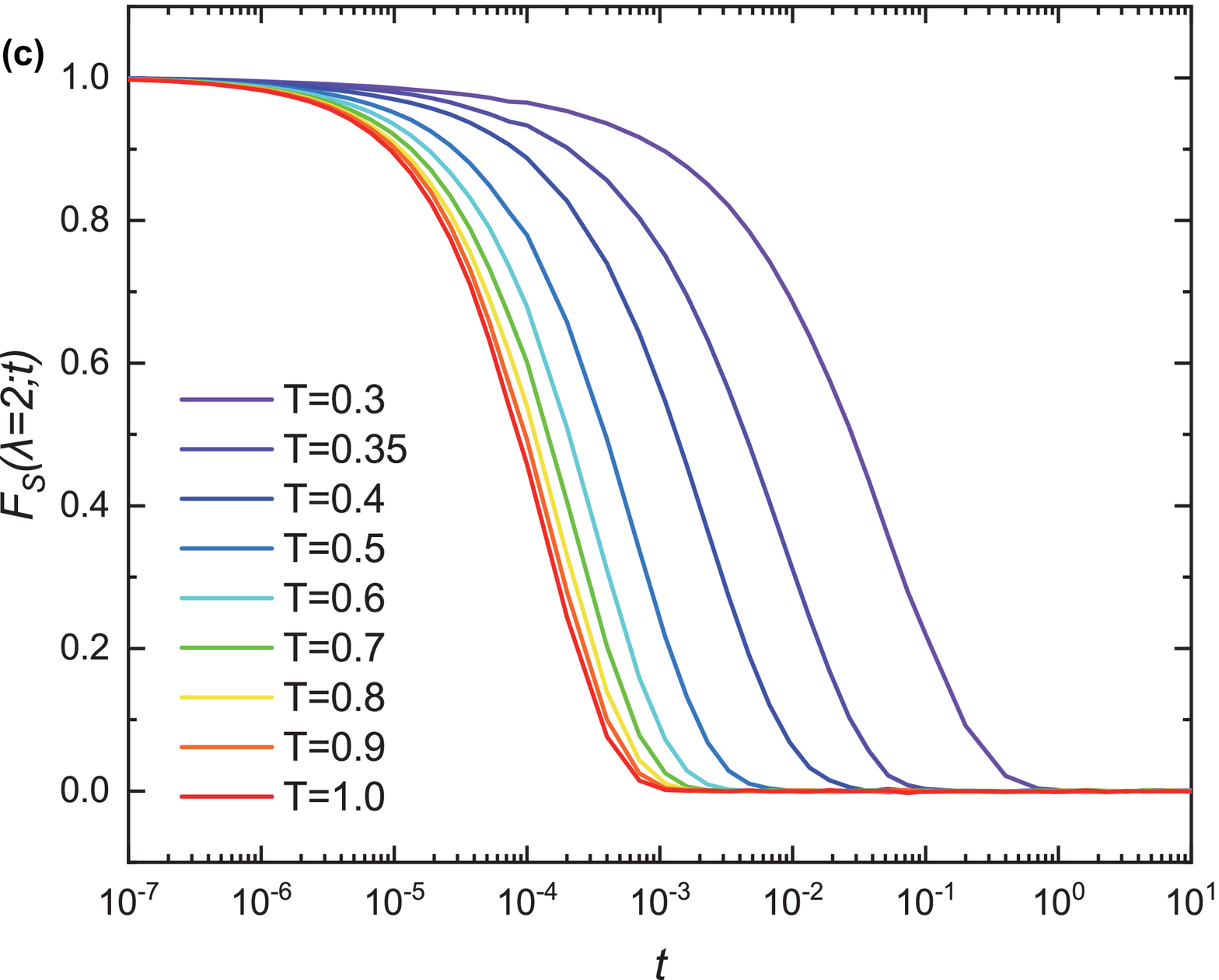}
\caption{Self-intermediate scattering function $F_s(\lambda;t)$ against time $t$ at wavelength $\lambda=2$ and various temperatures $T$ for 2D scalar MCLM with $M=10$ particle types (a), 3D scalar MCLM with $M=100$ particle types (b) and 2D vector MCLM with $M=25$ particle types (c). 
}
\label{SIF}
\end{figure}

\begin{figure}[tb]
\centering
\includegraphics[width=0.4\textwidth]{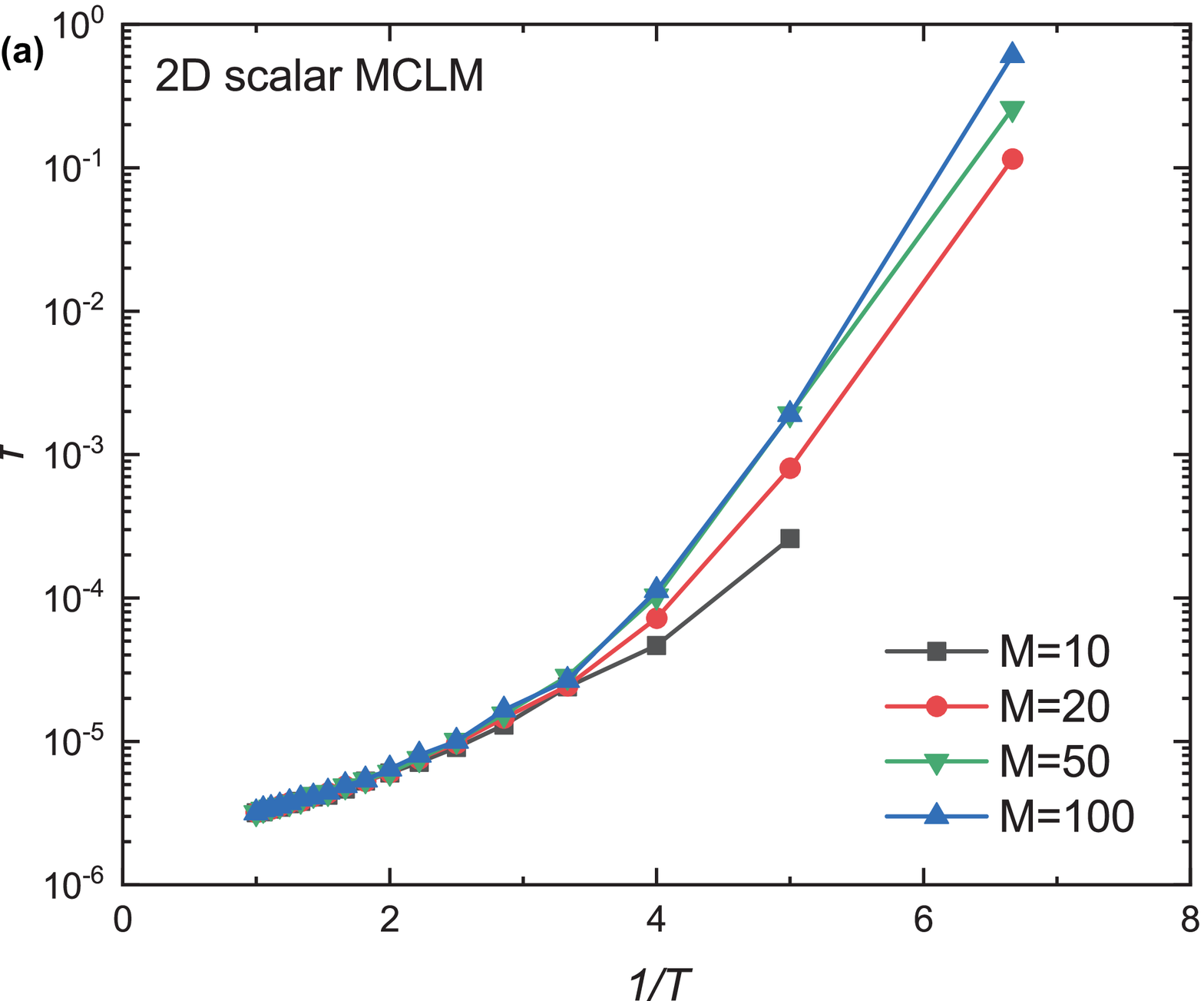}
\includegraphics[width=0.4\textwidth]{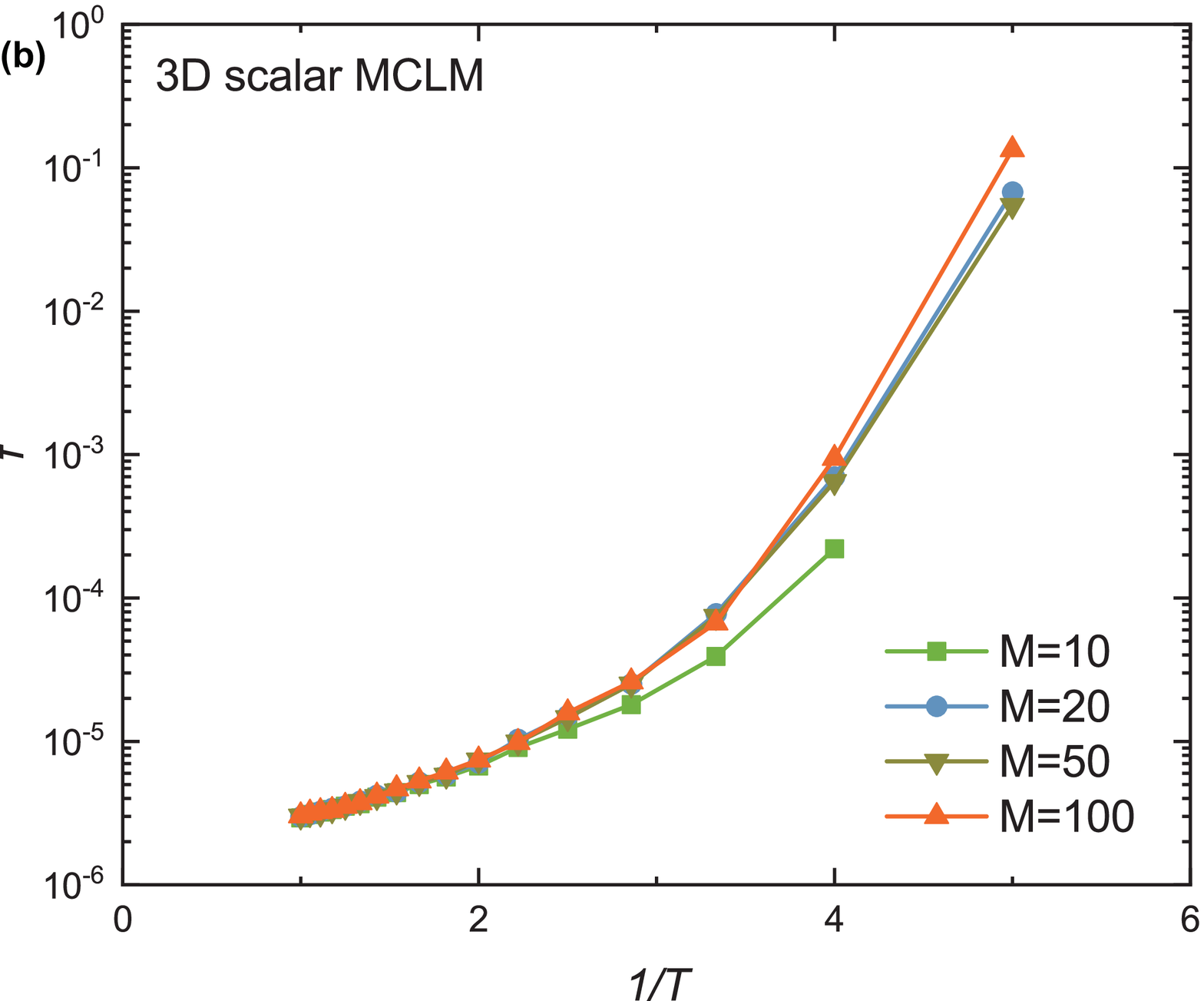}
\includegraphics[width=0.4\textwidth]{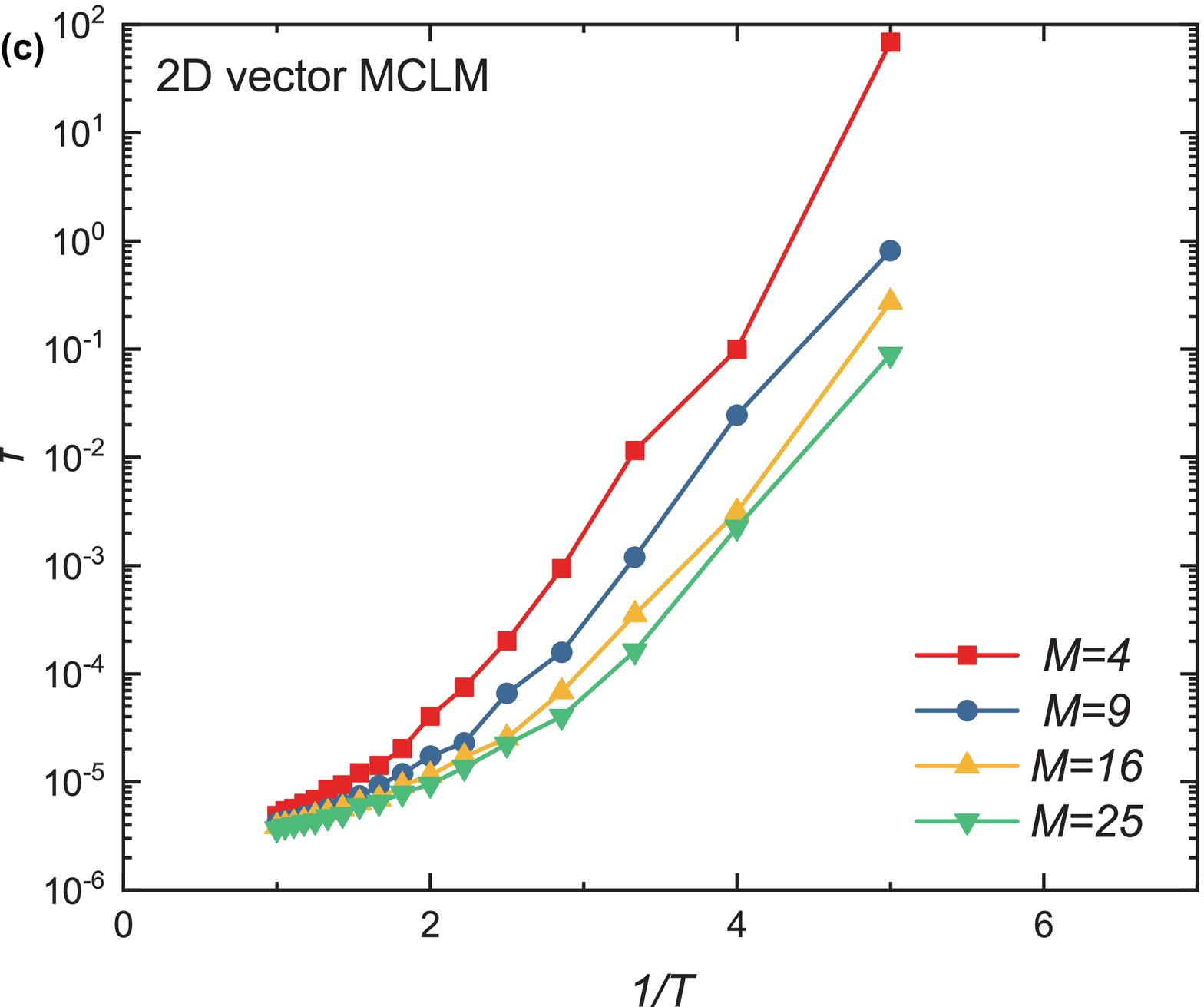}
\caption{Relaxation time $\tau$ against $1/T$ for 2D scalar (a), 3D scalar (b) and 2D vector (c) MCLM with $M$ particle types.
}
\label{TAU}
\end{figure}

\begin{figure}[tb]
\centering
\includegraphics[width=0.4\textwidth]{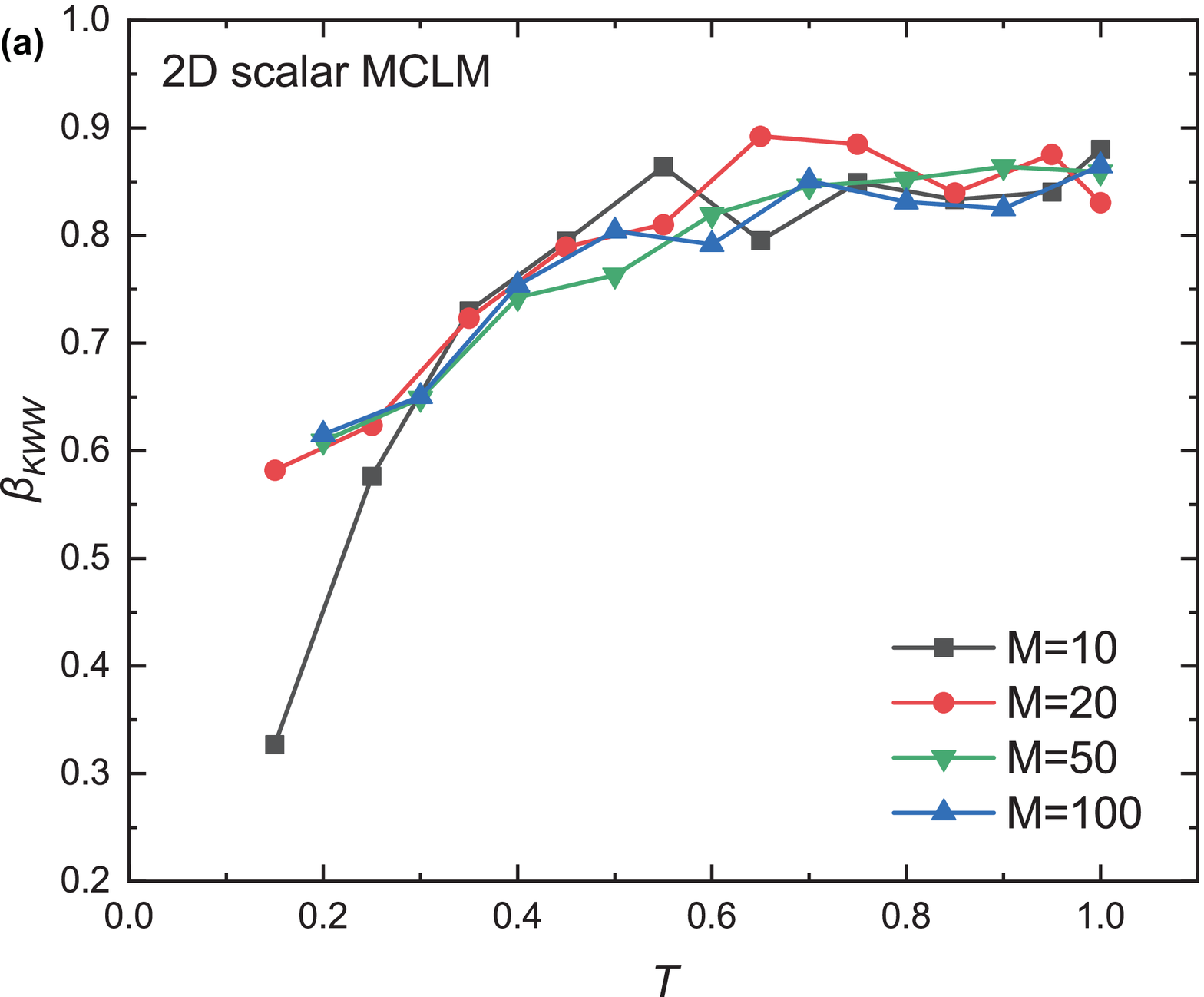}
\includegraphics[width=0.4\textwidth]{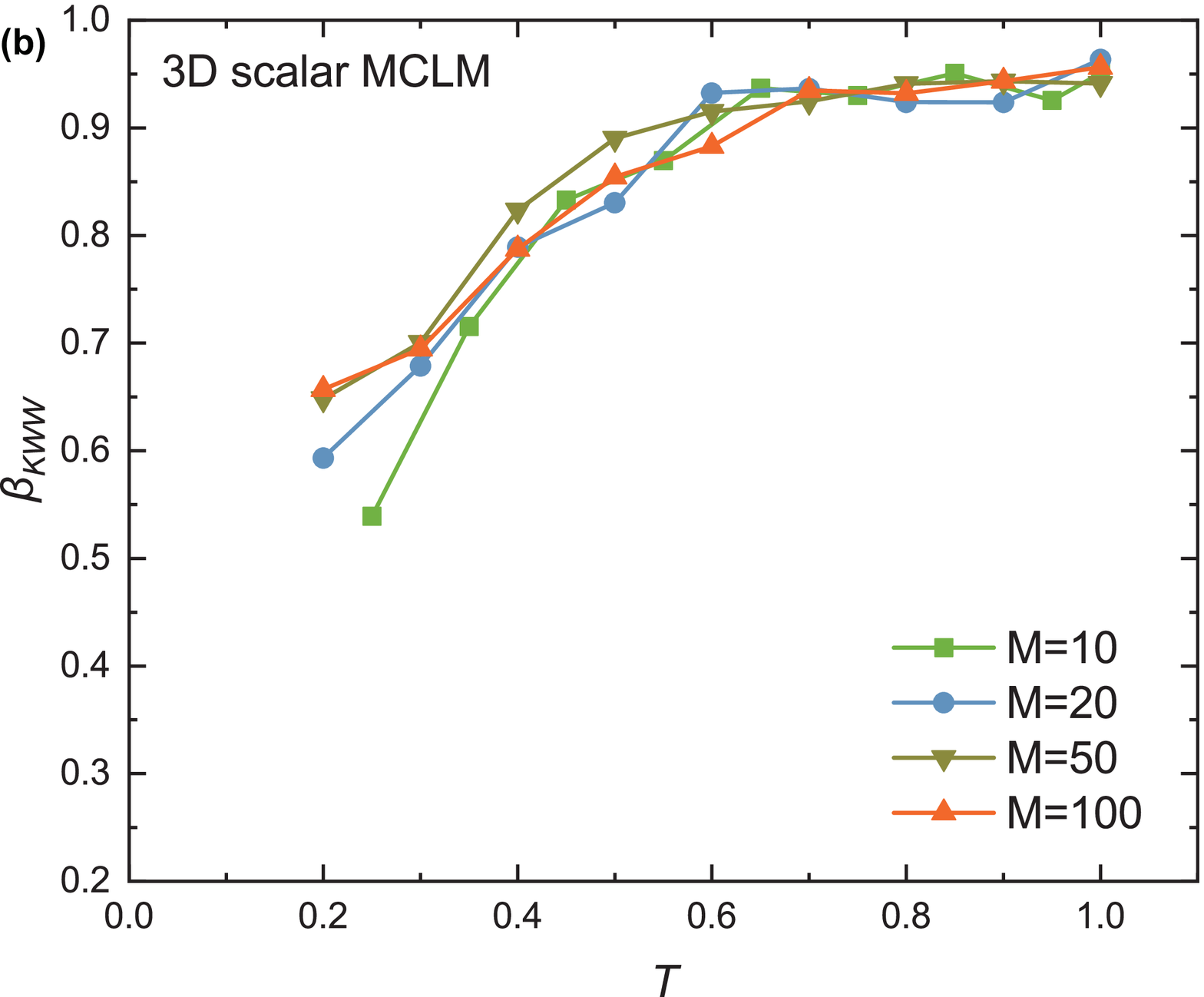}
\includegraphics[width=0.4\textwidth]{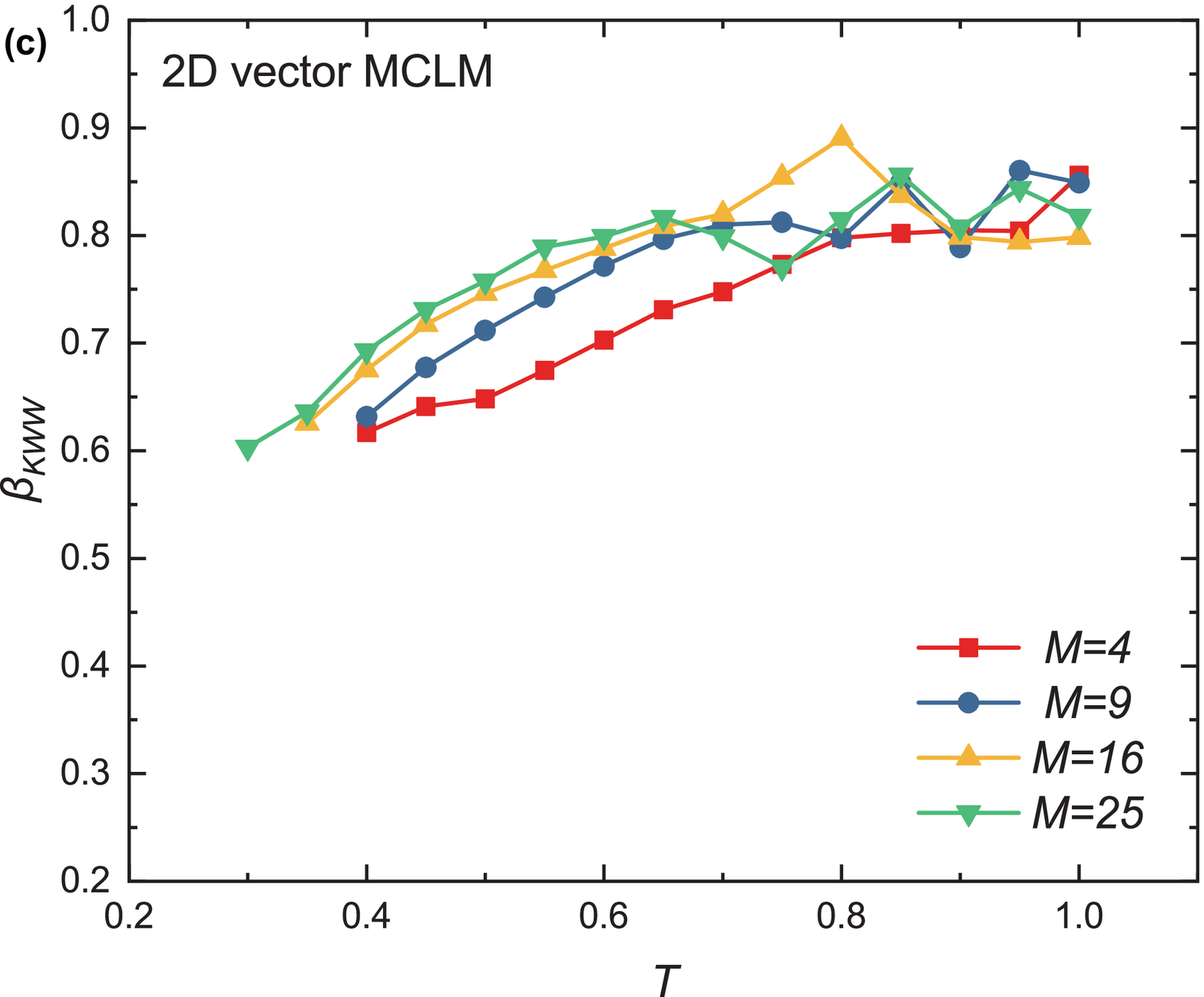}
\caption{Stretching exponent $\beta_{\text{KWW}}$ against $T$ for 2D scalar (a), 3D scalar (b) and 2D vector (c) MCLM with $M$ particle types.  
}
\label{BETA}
\end{figure}

From kinetic simulations, we measure the self-intermediate scattering function (SISF) $F_S$ defined as
\begin{equation}
F_S(\lambda;t)=\left\langle\exp[i\mathbf{q}\cdot(\mathbf{r}_l(t)-\mathbf{r}_l(0))]\right\rangle
\end{equation}
where the wavevector $q=|\mathbf{q}|=2\pi/\lambda$ with $\lambda=2$. \Fig{SIF} shows the results.
For typical glassy systems, a two-step decay of the SISF is expected.  In lattice models, the first decaying step is hardly discernible in the linear scale because of the lack of vibration \cite{zhang2017}. From the main decay observable in \fig{SIF}, the relaxation time $\tau$ is defined to be the time when $F_S$ drops to $1/e$. \Fig{TAU} plots values of $\tau$ obtained. Again, a super-Arrhenius temperature dependence is evident. 
In addition, the long-time functional form of the SISF is well approximated by the Kohlrausch-Williams-Watts form $A\exp(-(t/\tau)^{\beta_{\text{KWW}}})$, where $\beta_{\text{KWW}}$ is the stretching exponent while $A\simeq 1$ is a decay magnitude. Values of $\beta_{\text{KWW}}$ hence obtained are shown in \fig{BETA} and are found to decrease as $T$ decreases, similar to typical glasses. 

\subsection{Bond correlations}
Local static correlations, in general, exist in particle systems. To study correlations between the pair-interaction energies of neighboring bonds, we measure a normalized interaction correlation defined by
\begin{equation}
\rho= \frac{\mathrm{cov}(V_1,V_2)}{\mathrm{var}(V_1)}  
\label{rho}
\end{equation}
where '$\mathrm{cov}$' and '$\mathrm{var}$' denote covariance and variance and $V_1$ and $V_2$ are nearest neighboring interactions in perpendicular directions sharing one common particle.
As shown in Fig.~2 in the main text,
$\rho$ is non-zero and is negative in the 2D scalar MCLM for small $M$ at low $T$ so that annealed averaging approximation breaks down.
For the vector MCLM, the correlation vanishes under all conditions, supporting its exact thermodynamics for all $M$.

The correlation $\rho$ is quite moderate even when non-zero for the scalar MCLM. This is typical for systems without long-range correlations, as crystalization and segregation are suppressed by the direction dependence of the interaction $V_{\alpha k l}$.  
From Fig.~2 in the main text, we observe that the magnitude of $\rho$ seems to have saturated as $M$ is decreased from 20 to 10 at low $T$ for the scalar MCLM.
We find that for $M=20$, a simulation with a randomly selected set of interactions $V_{\alpha k l}$ can lead to either a positive or negative contribution to $\rho$. This may have resulted in a small $\rho$ after the ensemble averaging, For $M=20$, most individual simulations contribute negative correlations.

\begin{figure}[tb]
\centering
\includegraphics[width=0.47\textwidth]{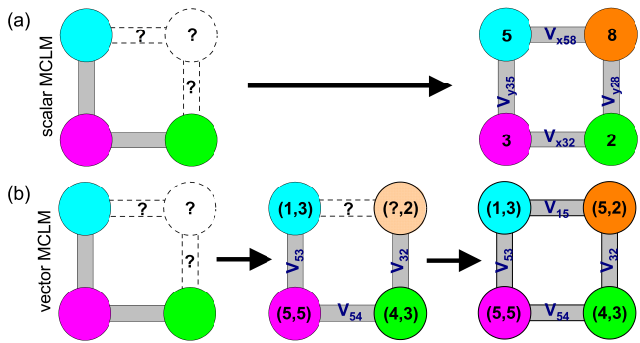}
\caption{
  (a) Schematic of a small $2\times 2$ scalar MCLM system with three particles (left panel). Assume that a fourth particle can be selected from a particle reservoir. This also determines the two remaining interactions (right panel). The selection of both interactions is thus a one-step process. At low $T$, a particle is an energetically favorable candidate if both interactions are of relatively low energies. 
  For small $M$, there will be few such candidates. The choice of the interaction in the $x$ direction implies the particle being selected and hence also on the interaction in the $y$ direction. This explains the correlation between the two neighboring interactions. 
  (b) For the vector MCLM, particle selection can effectively be broken down in a two-step process. To select the fourth particle (left panel), the $y$ type index can first be chosen  (mid panel). For any chosen $y$ type index, the $x$ type index can still take any value via an appropriate choice of the particle (right panel). Correlation between the interactions is therefore absent.
}
\label{explanation}
\end{figure}

The vector MCLM is introduced with correlations between bonds eliminated to achieve exact solvability even at low temperatures. 
\Fig{explanation} shows schematic diagrams explaining why bond correlations are present in the scalar MCLM but not in the vector MCLM. For the scalar MCLM at small $M$, few particle types can provide low-energy interactions at a site in all directions and the energy signatures, i.e. a characteristic set of pairs of coupled horizontal and vertical interactions, of such a small number of particles results in the correlations. At large $M$, all signatures average out. This explains why the DPLM with infinitely many particle types does not admit any interaction correlation.

\section{Approximations on Scalar MCLM}

We study the thermodynamic properties of the scalar MCLM using three approximate methods: the annealed averaging approximation and the uncorrelated particle approximation with or without excitations. All methods become exact at large $M$. The accuracy of the methods deteriorates as $T$ decreases due to stronger correlations. 
At low $T$, the uncorrelated particle approximation with excitation, which requires weaker assumptions and is most realistic, is expected to be more reliable. It predicts a positive specific entropy $S/N$ at all $T>0$ and a smooth turn before converging to 0 at $T=0$. This shows the best agreement with our simulations.

\subsection{Annealed averaging approximation}
\label{Sec:annealed}
\newcommand{\s}{{\{s_i\}}}
\newcommand{\Zbond}{Z_{\mathrm{bond}}}
\newcommand{\Vx}{V^x}
\newcommand{\Vy}{V^y}

The DPLM admits exact equilibrium statistics at all $T$ as have been derived  \cite{zhang2017} and verified by accurate numerical measurements of system energy \cite{zhang2017,lee2020} and pair-interaction distribution \cite{zhang2017,lulli2020}. The solvability follows from the result that averaging over particle permutations is equivalent to annealed averaging and gives exact statistics at all temperatures \cite{zhang2017}.

The MCLM reduces to the DPLM when the number of particle types $M$ is large. Annealed averaging, though no longer exact, can thus serve as an approximation, especially at large $M$.
Our calculations are closely related to those in \cite{zhang2017,lee2020} for the DPLM.

We consider a fully occupied lattice for simplicity. The canonical partition function is
\begin{equation}
    \label{Z}
    Z =  \sum_{\s} e^{-\upbeta E}
\end{equation}
where the sum is over all possible system states $\s$ and $\upbeta=1/k_BT$. Using \eq{app:systemEnergy}, we have
\begin{eqnarray}
  \label{Zn1}
    Z = \sum_{\s} \prod_{<ij>^*}  e^{-\upbeta V_{\alpha_{ij} s_i s_j}}  ~.
\end{eqnarray}
where the product is over all neighboring sites $i$ and $j$ with $j$ at the East or North of $i$. Applying an annealed averaging over independent variables $V_{\alpha  kl}$, we get
\begin{equation}
  Z =\frac{N!}{(N/M)!^M}  \roundbk{\Zbond} ^{~N_b}
\label{annealedZ}
\end{equation}
where $\Zbond$ is the partition function of a bond defined as
\begin{equation}
  \Zbond = \int\exp(-\beta V) g(V) \mathrm{d}V
\end{equation}
and $N_b=N d$ is the total number of interactions in the system in $d$ dimensions.
The prefactor $N!/(N/M)!^M$ comes from the number of arrangements of $M$ particle types each with $N/M$ particles on a lattice of $N$ sites. 
Taking the logarithm of \eq{annealedZ} gives:
\begin{equation}
\ln Z = N_b\ln \Zbond    +N\ln M
\label{logAnnealedZ}
\end{equation}
where  
Stirling's formula has been used.
Using the thermodynamic relation $F=E-TS$ with the free energy $F= -(1/\beta) \ln Z$, 
the entropy $S$ follows
\begin{equation}
\frac{S}{N}={ d~} \frac{\langle V \rangle-U }{T}+k_B\ln M
\label{entropy1}
\end{equation}
where $\langle V \rangle = - \partial \ln Z_{\text{bond}} / \partial \beta $ is the average pair interaction and $U = - (1/\beta) \ln \Zbond$ is the free energy of an interaction. 
Taking the  interaction distribution $g(V)$ as a uniform distribution in $[0,\Delta V]$ and after some algebra, \eq{entropy1} gives
 \begin{equation}
{\frac{S}{N}=k_B \bracebk{ d\squarebk{1+\frac{\beta\Delta V}{1-e^{\beta \Delta V}}+\ln\left(\frac{1-e^{-\beta \Delta V}}{\beta\Delta V}\right)}+\ln M }}
\label{entropy2}
\end{equation}
which has been applied to estimate $S/N$ as reported in Fig.~1.

To estimate the Kauzmann temperature, \eq{entropy2} is simplified at $k_BT<<1$ to 
\begin{equation}
  {\frac{S}{N}=k_B\left[d\left(1+\ln\left(\frac{k_B T}{\Delta V}\right)\right)+\ln M \right]}
  \label{S9}
\end{equation}
It is clear from \eq{S9} that $S$ becomes negative at small $T$. The temperature at which $S=0$ is the  Kauzmann temperature $T_K$ and is given in the annealed averaging approximation by
\begin{equation}
T_K=\frac{\Delta V}{e k_B M^{1/d}}.
\label{app:TK}
\end{equation}
\begin{figure}[tb]
\centering
\includegraphics[width=0.2\textwidth]{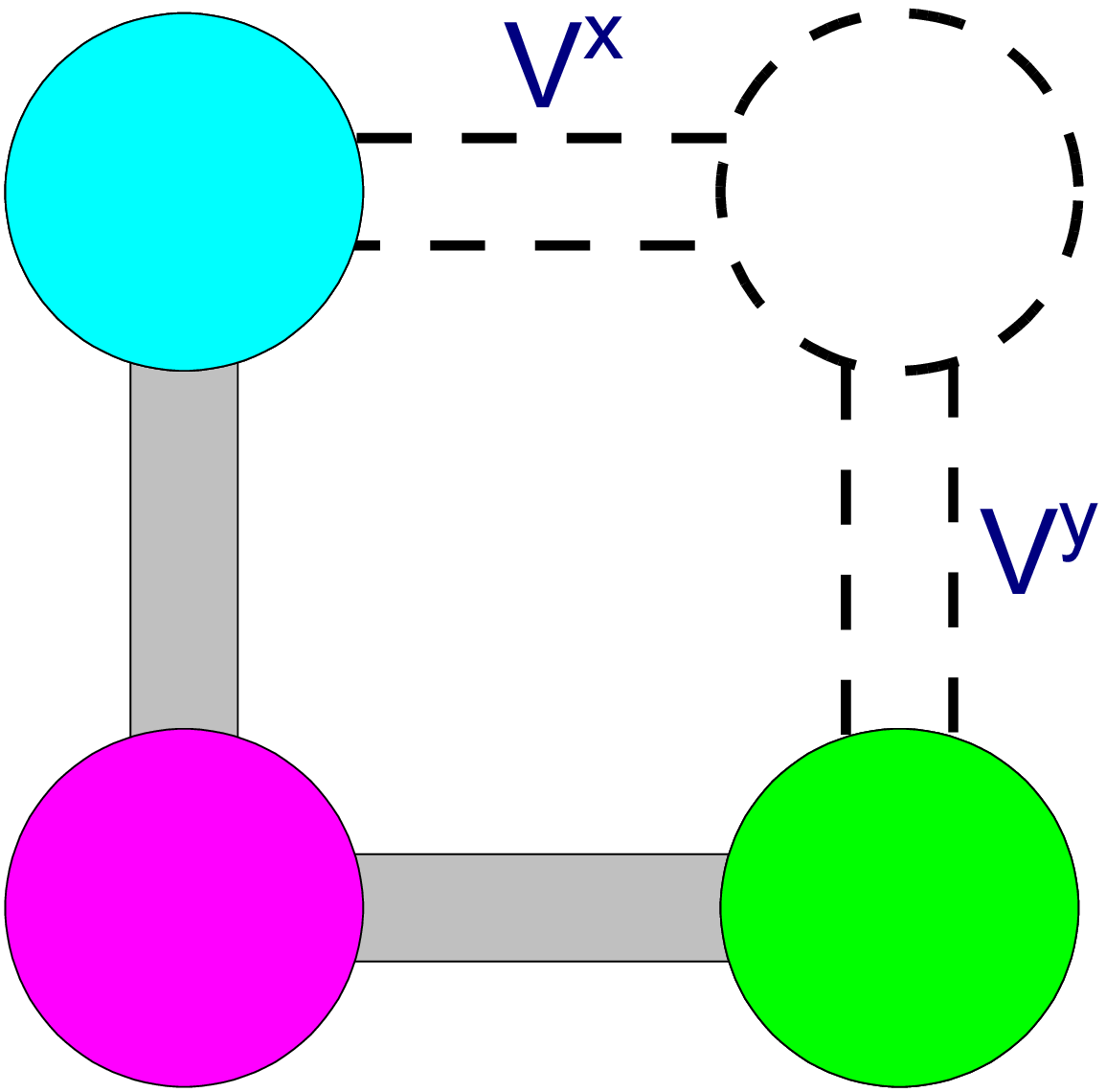}
\includegraphics[width=0.4\textwidth]{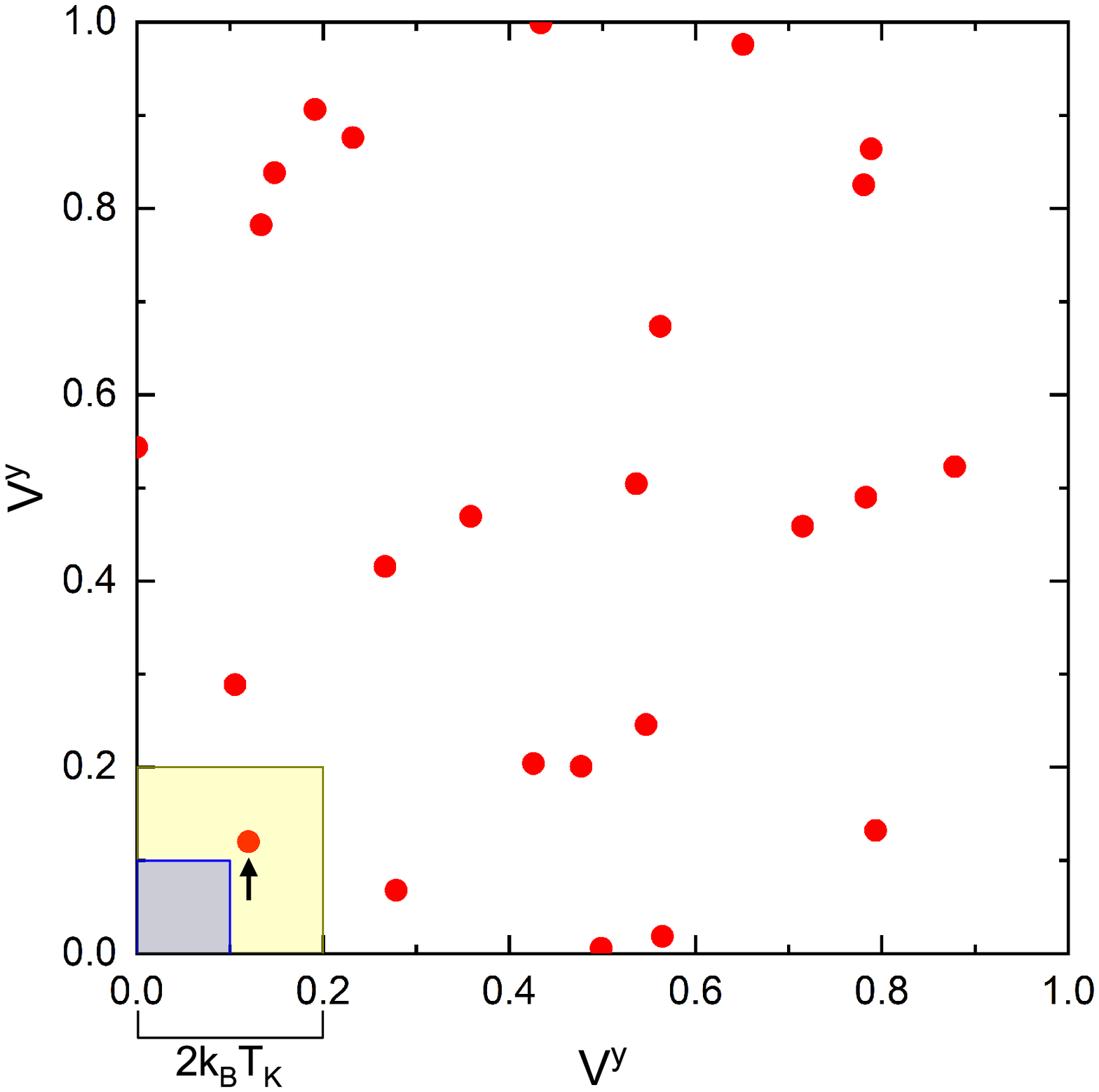}
\caption{
  Upper panel: A schematic showing that adding a particle to a kink site (dashed circle) adds two interactions $\Vx$ and $\Vy$ to the system. Lower panel: Consider a scalar MCLM with $M=25$ particle types. Possible interactions $\Vx$ and $\Vy$ are represented by $M$ red dots on the $\Vy$ against the $\Vx$ plot.
At $T_K$, typically one dot (arrow) resides within a box of width $2k_BT$ (yellow region).
For $T  < T_k$, the box shrinks (grey region). It then contains fewer than one dot on average but the number of the energetically favorable particle (arrow) remains one.
}
\label{VxVy}
\end{figure}

\subsection{Uncorrelated-particle approximation without excitation}
\label{section::noexcitation}

The annealed averaging approximation effectively assumes uncorrelated interactions, which also implies uncorrelated particle pairings.  A weaker assumption of only uncorrelated particles will be adopted now.
Consider constructing a configuration by inserting particles at the kink sites sequentially from the left to the right, repeated layer by layer from the bottom to the top of the lattice. 
We assume that the two particles around the kink site are uncorrelated
and can be of any type independently. As schematically illustrated in the upper panel in \fig{VxVy},
inserting a kink particle results in two new interactions, denoted by $\Vx$ and $\Vy$, which are thus independent of each other.
For any given configuration around the kink site, the pair $(\Vx, \Vy)$  takes one of $M$ possible choices corresponding to the $M$ particle types. They can be represented as $M$ random points uniformly distributed in a $\Vx$ versus $\Vy$ plot as illustrated in \fig{VxVy}.

A simple approximate derivation of \eq{app:TK} can now be explained.
Energetically favorable particle types are mainly those with both interactions $\Vx$ and $\Vy$ within $2k_BT$ (yellow area in \fig{VxVy}), accounting for a mean and fluctuation both equal to $k_BT$. The average number of favorable types is then
\begin{equation}
  M_{kink} = 4 M \roundbk{\frac{k_B T}{\Delta V}}^2.
  \label{Mkink}
\end{equation}
The specific entropy can be estimated by $S/N = k_B \ln M_{kink}$.
Then, $T_K$ is defined as the temperature at which $M_{kink} = 1$ so that $S/N = 0$.
After a generalization to $d$ dimensions, this leads to
\begin{equation}
  T_K = \frac{\Delta V}{2 k_BM^{1/d}}
  \label{TKsimple}
\end{equation}
which differs slightly from \eq{app:TK} only in the coefficient.
For $T \agt T_K$, any of the $M_{kink} \agt 1$   particle types are energetically favorable and can be added to the kink site, resulting in a positive $S/N$ which decreases rapidly as $T$ decreases.

For $T < T_k$ in contrast, \eq{Mkink} gives $M_{kink} < 1$, which can lead to an apparent  negative entropy. However, $M_{kink} < 1$ simply reflects the breakdown of the assumption that both interactions must be below $2k_BT$. In fact, at very low $T$, the type providing the smallest interactions will be favored (arrow in \fig{VxVy})), no matter the actual values. Therefore, \eq{Mkink} should be amended to
\begin{equation}
  M_{kink} = \max\bracebk{4 M \roundbk{\frac{k_B T}{\Delta V}}^2, 1},
  \label{Mkink2}
\end{equation}
to avoid any unphysical negative entropy. Note that the sharp turn of \eq{Mkink2} at $T_K$ will be smoothed in a more careful calculation considering also excitations to higher-energy interactions, as explained in the next section. 

From the derivation of \eq{TKsimple}, the dimensional dependence of $T_K$ can be intuitively understood as follows. At $T_K$, there must be one energetically favorable particle type that provides $d$ energetically favorable interactions, i.e. those within $2k_BT$, in each of the $d$ directions with the $d$ neighbors. At higher dimension $d$, the constraints become more numerous and harder to satisfy. The box size $(2k_BT_K)^d$ in the $(V^x, V^y, \dots )$ space must increases to include at least one particle type. Hence, $T_K$  increases.

\subsection{Uncorrelated-particle approximation with excitations}
\label{section:uncoptcle}
\begin{figure}[tb]
\centering
\includegraphics[width=0.4\textwidth]{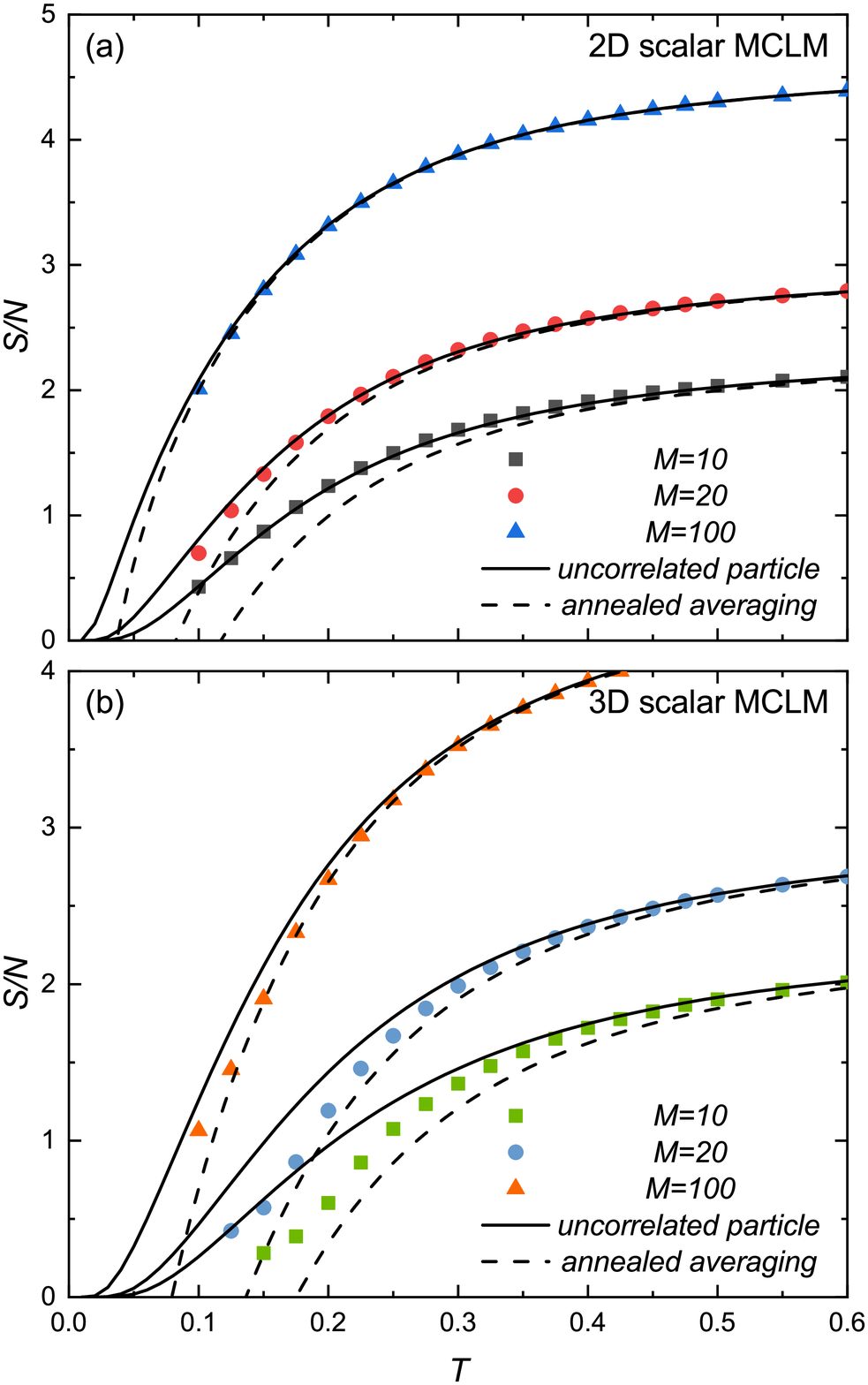}
\caption{
  (a) Entropy per particle $S/N$ against temperature $T$ for scalar MCLM with $M$ particle types in 2D (a) and 3D (b) using the same data from Fig.~1 in the main text.
  Results are compared with uncorrelated particle approximation (solid lines) and annealed averaging approximation (dashed lines).  
}
\label{Vakl_uncoPtcle}
\end{figure}

\newcommand{\Vkink}{V_{\text{ptcle}}}
\newcommand{\Vkinkg}{V_{\text{ptcle},1}}
\newcommand{\Gd}{G_{\text{dis}}}

In \app{section::noexcitation}, the interactions $\Vx$ and $\Vy$ of a kink particle are assumed either within $2k_BT$ or of the lowest possible energies of the $M$ particle types. We now perform a more detailed calculation including excitations to higher energy interactions using standard thermodynamic methods. The calculation is closely analogous to that in \app{Sec:annealed}.

Noting that $\Vx$ and $\Vy$ follow the  distribution $g(V)$,
the total energy ${\Vkink=\Vx+\Vy}$ of the kink particle then follows the   distribution
\begin{equation}
G(\Vkink)=g(\Vx)\circ g(\Vx)
\label{G}
\end{equation}
where '$\circ$' denotes convolution and $g(V)$ is uniform in $[0, \Delta V]$ with $\Delta V=1$.

The continuum distribution $G(\Vkink)$ accounts for all possible values of $\Vkink$ for arbitrary neighbors at the kink site. For a given pair of neighbors, there are only $M$ possible values of $\Vkink$  corresponding to the $M$ possible types of the kink particle.
Then, $\Vkink$ follow
 a discrete distribution 
\begin{equation}
\Gd(\Vkink)=\frac{1}{M}\sum_{\mu=1}^{M}\delta(\Vkink-\epsilon_\mu).
\label{GDiscrete}
\end{equation}
where $\epsilon_\mu$ denotes the $M$ possible values of $\Vkink$.
Averaging $\Gd(\Vkink)$ over all possible neighbors at the kink site should restore $G(\Vkink)$.
Rather than performing such an average, a simpler approach is to consider a single typical sample set of $\epsilon_\mu$. 
Analogous to standard techniques for generating non-uniform random numbers using cumulative probability, 
we solve the typical $\epsilon_\mu$ from 
\begin{equation}
\mathcal{G}_c(\epsilon_\mu)=\frac{1}{M}(\mu-1/2).
\end{equation}
where $\mathcal{G}_c(\Vkink)$ is the cumulative distribution function of $G(\Vkink)$ defined as
\begin{equation}
\mathcal{G}_c(\Vkink) = \int_{-\infty}^{\Vkink} G(V) \mathrm{d}V.
\end{equation}
Using this approach, $\Gd(\Vkink)$ converges to $G(\Vkink)$ at large $M$.

Invoking again the uncorrelated-particle assumption, the system partition function can be expressed in the factorized form, analogous to \eq{annealedZ},
\begin{equation}
Z=M^{N}( Z_{\text{pctle}} )^{N}
\label{ZUncorrPtcle}
\end{equation}
where the partition function of a particle is
\begin{equation}
Z_{\text{pctle}}  =\frac{1}{M}\sum_{\mu=1}^{M}\exp(-\beta \epsilon_\mu).
\end{equation}
A factor $1/M$ is introduced in the definition of $Z_{\text{pctle}}$ so that notations here can be more consistent with those in \app{Sec:annealed}. 
The
entropy $S$ can be obtained from
\begin{equation}
\frac{S}{N}=\frac{\langle \Vkink \rangle - U_{\text{ptcle}}}{T}+k_B\ln M
\end{equation}
where $\langle \Vkink \rangle = - \partial \ln Z_{\text{ptcle}} / \partial \beta  $ is the average particle energy and  $U_{\text{ptcle}} = - (1/\beta) Z_{\text{pctle}}$ denotes the free energy of a particle.

\Fig{Vakl_uncoPtcle} shows $S/N$ hence calculated. Good agreement with simulation results is observed at large $M$ and high $T$. More importantly, a positive $S/N$ is now ensured and it decreases to 0 smoothly at $T=0$.

\section{Exact thermodynamics of Vector MCLM}
\begin{figure}[tb]
\centering
\includegraphics[width=0.4\textwidth]{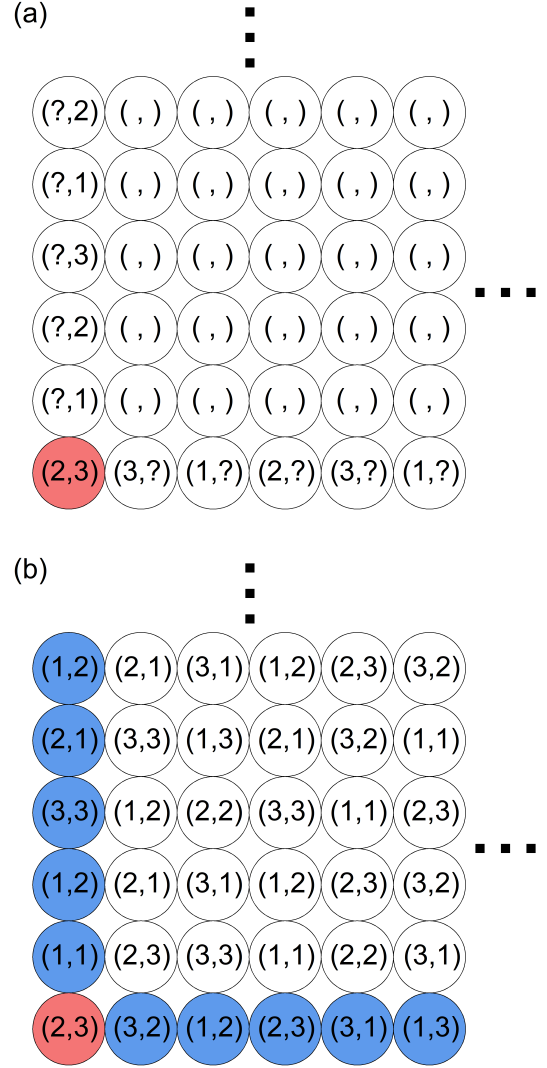}
\caption{
  A schematic of the construction of a ground state of the vector MCLM with $M=9$ so that $\mathcal{M}=3$. (a) Once the particle type at the bottom left corner (red) is selected, the $x$ type indices in the bottom row and the $y$ type indices in the leftmost column are determined. (b) We then select the $y$ type indices in the bottom row and the $x$ type indices in the leftmost column (blue). Particle types of the whole lattice are now completely fixed. 
}
\label{vectorMCLMschematic}
\end{figure}

\subsection{Ground states}
\label{section::groundstate}

An example of the ground states of a small system is already shown in Fig.~4 in the main text. 
The system energy $E$ is zero with all interactions zero.
Such ground states are highly degenerate. In the thermodynamic limit, we can calculate the degeneracy as will now be explained. 
\Fig{vectorMCLMschematic}(a) shows schematically part of a large $L \times L$ system with $M=\mathcal{M}^2=9$ particle types. To construct a system configuration, we first select any of the $M$  particle types at the bottom left corner site, leading to a multiplicity of $M$. In a ground state, all interactions, as given by \eq{app:Vkala}, must be 0. The type indices $s_{i}(x)$ at all sites $i$ at the bottom row are then uniquely determined. However,  the other indices $s_{i}(y)$ (question marks in \fig{vectorMCLMschematic}(a)) are free to take any of the $\mathcal{M}$ possible values, contributing to a multiplicity of $\mathcal{M}^{L-1}$. A similar argument can be applied to every site on the leftmost column. Once particle types on these boundary sites are chosen, particle types on the whole lattice are uniquely determined (\fig{vectorMCLMschematic}(b)). These result in a total degeneracy 
$\Omega=\mathcal{M}^{2L}=M^L=M^{\sqrt{N}}$.
The residue entropy $S_0=k_B\ln \Omega$  at $T=0$ is thus given by 
\begin{equation}
  S_0 = \sqrt{N} k_B \ln M  . 
\end{equation}
Therefore, $S_0$ diverges at large $N$. However, it is non-extensive. The residue entropy per particle is $S_0/N = k_B \ln M /\sqrt{N}$ which vanishes at large $N$. 

Note that the above argument holds exactly under open boundary conditions or periodic boundary conditions if $L$ is a multiple of $\mathcal{M}$. Otherwise, a vanishing density of defects may occur, which does not alter our results qualitatively. In addition, as we assume an equal number of particles $N/M$ of each type in the system, the ground states constructed using the method above fulfill these constraints only on average.
Strict fulfillment of the constraints decreases $S_0$ slightly but has a negligible impact on the statistics of individual particles in the thermodynamic limit. 
At small $N$, a direct construction of the configurations under the constraints is non-trivial. We have thus obtained the example of the ground state in Fig.~4 from a simulation instead. 

\subsection{Exact thermodynamics}
\label{section::exactthermodynamics}
Here, we explain the exact analysis of a fully occupied 2D vector MCLM. Generalizations to a finite void density and higher dimensions are straightforward. On a $L \times L$ lattice, denote site $i$ by 
its lattice coordinates $(m,n)$, where $m,n = 1, 2, \dots, L$.
Then, the particle type $\mathbf{s}_i$ at site $i$ can equivalently be denoted as
$\mathbf{s}_{mn} = ( s_{mn}(x), s_{mn}(y))$.
To simplify the notation, interactions to the East and North of site $(m,n)$ are denoted by
$V_{mn}^x$ and $V_{mn}^y$, i.e.
\begin{equation}
\begin{split}
  V_{mn}^x&=V_{s_{mn}(x) s_{(m+1)n}(x)}\\
  V_{mn}^y&=V_{s_{mn}(y) s_{m(n+1)}(y)}.  
\end{split}
\end{equation}
Then, the system energy $E$ in \eq{app:Evector} can be recast into
\begin{equation}
\begin{split}
E=E_x+E_y\\
\end{split}
\end{equation}
where $E_x$ and $E_y$ are the total interaction energy in $x$ and $y$ directions and are given by
\begin{align}
E_x&=\sum_{m,n}V_{mn}^x\label{Ex}\\
E_y&=\sum_{m,n}V_{mn}^y.
\end{align}
The partition function $Z$ of the system is
\begin{equation}
\begin{split}
Z&=\sum_{\{\mathbf{s}_{mn}\}}\exp(-\beta E)\\
\end{split}
\end{equation}
where the sum is over the set of all configurations ${\{\mathbf{s}_{mn}\}} \equiv \{(s_{mn}^x,s_{mn}^y)\}$. 
By noting that $E_x$ and $E_y$ depend only on $s_{mn}^x$ and $s_{mn}^y$ respectively, 
$Z$ can be factorized to
\begin{equation}
\begin{split}
Z&=Z_x Z_y\\
\end{split}
\end{equation}
where
\begin{align}
  Z_x&=\sum_{\{s_{mn}^x\}}\exp(-\beta E_x)\label{Zx}\\
  Z_y&=\sum_{\{s_{mn}^y\}}\exp(-\beta E_y).
\end{align}

We first evaluate $Z_x$.
Consider a particle of a given type and its neighbor on the East. 
According to \eq{app:Vkala}, there is a one-one correspondence between the $x$ type index of its neighbor and the value of the interaction.  
Generalizing to all sites at row $n$, there is also a one-one correspondence between the sets ${\{s_{mn}^x\}}$ and ${\{V_{mn}^x\}}$ with $m=2,3, \dots, L$ for any given ${\{s_{1n}^x\}}$ at the boundary. 
Using also \eq{Ex}, \eq{Zx}
thus becomes 
\begin{equation}
\begin{split}
  Z_x&=
  \mathcal{M}^L
  \sum_{\{V_{mn}^x\}}\exp\left(-\beta \sum_{m,n}V_{mn}^x\right)\\
\end{split}
\end{equation} 
  where the sum is now over all possible values of the interactions consistent with the periodic boundary conditions and
the factor {$\mathcal{M}^L$}
accounts for all possible ${\{s_{1n}^x\}}$. 
At the thermodynamic limit, the factor {$\mathcal{M}^L$} and the constraints on $V^x_{mn}$ imposed by the periodic boundary conditions have a vanishing impact on the statistics of individual particles and are both neglected. We hence obtain the factorized form
\begin{equation}
Z_x={\mathcal{M}^N} \prod_{m,n}Z_{bond} 
\label{factorizedZx}
\end{equation}
where $Z_{bond}$ is the partition function of an interaction defined by
\begin{equation}
  Z_{bond} = \frac{1}{\mathcal{M}}\sum_{\mu=1}^{\mathcal M}\exp(-\beta V_\mu)
  \label{factorizedZx2}
\end{equation}
  with the set of possible interactions $V_\mu$ given in \eq{Vvector}. A factor $1/\mathcal{M}$ is introduced in the definition of $Z_{bond}$ so that notations here can be more consistent with those in \app{Sec:annealed}.

By symmetry, $Z_x=Z_y$. Similar to calculations in Sec. \app{Sec:annealed} and noting that $M = \mathcal{M}^d$ in $d$ dimensions, we get
\begin{equation}
\ln Z = N_b\ln \Zbond    +N\ln M
\label{logAnnealedZ2}
\end{equation}
and
\begin{equation}
\frac{S}{N}={ d~} \frac{\langle V \rangle-U }{T}+k_B\ln M
\label{entropyU}
\end{equation}
where $\langle V \rangle = - \partial \ln Z_{\text{bond}} / \partial \beta $ and $U = - (1/\beta) \ln \Zbond$. \Eq{entropyU} has been applied to calculate the exact result in Fig.~3(a) in the main text.

\subsection{Kauzmann paradox for the harmonic oscillator?}
\label{section::harmonicoscillator}

\begin{figure}[htbp]
\centering
\includegraphics[width=0.4\textwidth]{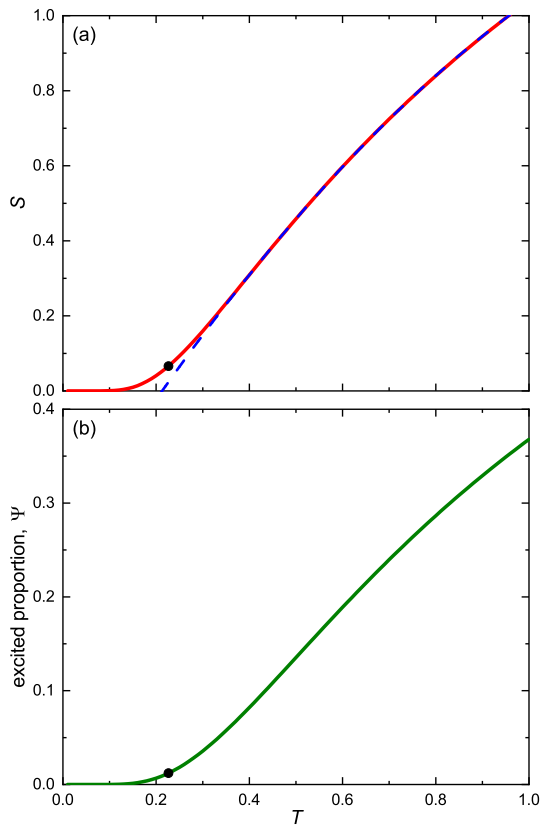}
\caption{
  Equilibrium properties of a quantum harmonic oscillator in one dimension, showing entropy $S$ (a) and excited proportion (b) against temperature $T$.
  }
\label{harmonic}
\end{figure}

We consider a quantum harmonic oscillator in one-dimension using dimensionless units with 
angular frequency $\omega=1$, mass $m=1$,  Planck's constant $\hbar=1$, and Boltzmann constant $k_B=1$. It is straightforward to show that the entropy is exactly given by
\begin{equation}
  S=-\ln(1-e^{-\beta})+\beta\frac{e^{-\beta}}{1-e^{-\beta}}
\end{equation}
where $\beta = 1/ k_B T$.
\Fig{harmonic}(a) shows $S$ plotted against $T$. We observe that as $T$ decreases, $S$ first decreases steadily. A naive extrapolation  (dashed line) of the high-temperature data can give a finite temperature $T_K$ at which the entropy seems to vanish. This is analogous to the Kauzmann paradox.
From the exact result, before reaching $S=0$, a smooth turn in fact occurs so that $S$ vanishes only at $T=0$. 

\Fig{harmonic}(b) plots the excited proportion $\Psi$ in an ensemble of independent oscillators which equals the probability of excitation beyond the ground state. It is straightforward to show that 
\begin{equation}
\Psi = 1-2e^{-\beta/2}\sinh{(\beta/2)}.
\end{equation}
Defining the turn of the entropy at $T=T_{0.7}$ (black circles in \fig{harmonic}(a) and (b)) at which the slope of $S$ has decreased to 70\% of its maximum value. The excited proportion at $T_{0.7}$ is {$\Psi=1.2\%$}. This small value of $\Psi$, which is of the same order of magnitude obtained for the vector MCLM as explained in the main text, shows again that the turn in the entropy occurs when the system is very close to the ground state. 

For the harmonic oscillator, the entropy turns rather suddenly and very close to the ground state. 
This would not lead to a paradox as it is exactly solvable and the turn is easily understood. However, it illustrates thoroughly how the naive extrapolation fails. 
The phenomenon is fully analogous to that in the vector MCLM. The similarity is not only qualitative. For the uniform discrete interaction distribution in \eq{app:Vkala} adopted for the vector MCLM, the energy levels are essentially similar to those in the harmonic oscillator except that they are bounded above by $\Delta V=1$. Results for the MCLM are thus quantitatively equivalent to that in the harmonic oscillator for $T \ll \Delta V$.
We believe that such a failure in simple extrapolation and a turn in the entropy may also apply to glass.

%

\bibliographystyle{apstest}
\end{document}